\documentclass[pmlr,twocolumn,10pt]{jmlr} 

\usepackage{times}  
\usepackage{helvet}  
\usepackage{courier}  

\usepackage{algorithm}
\usepackage{algpseudocode}

\usepackage{graphicx}
\usepackage{textcomp}
\usepackage{sidecap}

\usepackage{graphicx,xfp}
\usepackage{multicol}
\usepackage{multirow}
\usepackage{color}
\usepackage{xcolor}
\usepackage{booktabs}
\usepackage{graphicx}
\usepackage{nicefrac}
\usepackage{amsmath}
\usepackage{amssymb}
\usepackage{pifont}
\newcommand{\cmark}{\ding{51}}%
\newcommand{\xmark}{\ding{55}}%

\usepackage{array}
\newcolumntype{P}[1]{>{\centering\arraybackslash}p{#1}}

\usepackage{graphicx}
\usepackage{fancyhdr}
\usepackage{amssymb}
\usepackage{url}
\usepackage{hyperref}
\usepackage{color}
\usepackage{tikz}
\usetikzlibrary{positioning}
\usepackage{pgfplots}
\usepackage{marvosym}

\usepackage{float}
\usepackage{todonotes}

\usepackage{booktabs}

\usepackage{tikz}
\usepackage{tikzscale}
\usepackage{xspace}
\usepackage{pgfplots}
\usepgfplotslibrary{groupplots}
\pgfplotsset{compat=1.16}
\usepackage{xparse}
\NewDocumentCommand{\Log}{o}{%
  \IfNoValueTF{#1}{}{{}^{#1}\!}\log}%

\usepackage{xcolor}
\def\BibTeX{{\rm B\kern-.05em{\sc i\kern-.025em b}\kern-.08em
    T\kern-.1667em\lower.7ex\hbox{E}\kern-.125emX}}

\usepackage{longtable}
\usepackage{array}
\newcolumntype{L}[1]{>{\raggedright\let\newline\\\arraybackslash\hspace{0pt}}m{#1}}
\newcolumntype{C}[1]{>{\centering\let\newline\\\arraybackslash\hspace{0pt}}m{#1}}
\newcolumntype{R}[1]{>{\raggedleft\let\newline\\\arraybackslash\hspace{0pt}}m{#1}}

\jmlrvolume{LEAVE UNSET}
\jmlryear{2023}
\jmlrsubmitted{LEAVE UNSET}
\jmlrpublished{LEAVE UNSET}
\jmlrworkshop{Conference on Health, Inference, and Learning (CHIL) 2023} 

\title{Virus2Vec: Viral Sequence Classification Using Machine
  Learning}

\author{%
 \Name{Sarwan Ali} \hfill \Email{\textnormal{sali85@student.gsu.edu}} \\
 \addr{Georgia State University, Atlanta, USA}
 \AND
  \Name{Babatunde Bello} \hfill \Email{\textnormal{bbello1@student.gsu.edu} }\\
 \addr Georgia State University, Atlanta, USA 
 \AND
 \Name{Prakash Chourasia} \hfill \Email{\textnormal{pchourasia1@student.gsu.edu}} \\
 \addr Georgia State University, Atlanta, USA 
 \AND
  \Name{Ria Thazhe Punathil} \hfill \Email{\textnormal{rthazhepunathil1@student.gsu.edu}} \\
 \addr Georgia State University, Atlanta, USA 
 \AND
  \Name{Pin-Yu Chen} \hfill \Email{\textnormal{pin-yu.chen@ibm.com}} \\
 \addr  IBM Research, NY, USA 
 \AND
 \Name{Imdad Ullah Khan} \hfill \Email{\textnormal{imdad.khan@lums.edu.pk}}\\
 \addr  Lahore University of Management Sciences, Lahore, Pakistan 
 \AND
 \Name{Murray Patterson} \hfill \Email{\textnormal{mpatterson30@gsu.edu}} \\
 \addr Georgia State University, Atlanta, USA 
}

\begin{document}

\maketitle
\begin{abstract}
Understanding the host-specificity of different families of viruses
sheds light on the origin of, e.g., SARS-CoV-2, rabies, and other such
zoonotic pathogens in humans. It enables epidemiologists, medical
professionals, and policymakers to curb existing epidemics and prevent
future ones promptly. In the family Coronaviridae (of which SARS-CoV-2
is a member), it is well-known that the spike protein is the point of
contact between the virus and the host cell membrane. On the other
hand, the two traditional mammalian orders, Carnivora (carnivores) and
Chiroptera (bats) are recognized to be responsible for maintaining and
spreading the Rabies Lyssavirus (RABV). We propose Virus2Vec, a
feature-vector representation for viral (nucleotide or amino acid)
sequences that enable vector-space-based machine learning models to
identify viral hosts. Virus2Vec generates numerical feature vectors
for unaligned sequences, allowing us to forego the computationally
expensive sequence alignment step from the pipeline. Virus2Vec
leverages the power of both the \emph{minimizer} and position weight
matrix (PWM) to generate compact feature vectors. Using several
classifiers, we empirically evaluate Virus2Vec on real-world spike
sequences of Coronaviridae and rabies virus sequence data to predict
the host (identifying the reservoirs of infection). Our results
demonstrate that Virus2Vec outperforms the predictive accuracies of
baseline and state-of-the-art methods.
\end{abstract}

\paragraph*{Data and Code Availability}
We extracted the labeled Spike protein sequences for COVID-19 hosts dataset from GISAID~\footnote{\url{https://www.gisaid.org/}} and labeled Nucleotide genome sequences for rabies virus hosts dataset from RABV-GLUE~\footnote{\url{http://rabv-glue.cvr.gla.ac.uk/\#/home}}. Where the label is the name of the host for which we are classifying the sequences. 
Our preprocessed dataset and code are available online~\footnote{\url{https://github.com/sarwanpasha/Virus2Vec}}.

\paragraph*{Institutional Review Board (IRB)}
This work does not require IRB approval.


\section{Introduction}
The global COVID-19 pandemic has drawn the attention of researchers to
understand the origin of (zoonotic) viruses in humans.  In the case of
the coronaviruses (the family Coronaviridae), it has been established
that SARS was transmitted to humans from civets and MERS-CoV from
dromedary camels~\citep{reusken-2014-mers}. In contrast, it is widely
thought that SARS-CoV-2 (which causes COVID-19) originated from
bats~\citep{zhou_2020_pneumonia}. However, numerous zoonotic diseases
have been around for a while, and medical professionals have been
attempting to combat them better. The rabies virus is one such illness
with a near 100\% death rate after symptoms
appear~\citep{taylor2015global}. All mammal species are susceptible to
rabies. However, domestic dog bites account for up to 99\% of human
rabies cases~\citep{WHO_Rabies}. There is frequent spillover from dogs
into other carnivores, but typically this only results in transient
chains of transmission. Therefore, it is crucial to locate and keep an
eye on any potential wildlife reservoirs~\citep{worsley2020using}.  
It is crucial to understand the origins of such diseases to
create effective prevention and mitigation measures as well
as vaccines and therapeutics.
Pathogen sequence data are readily available, and genomic monitoring
is being used more and more frequently. To account for this, genomic
tools and classification algorithms need to be updated. 
New genomic technologies~\citep{gigante2020portable},
machine learning and learning-based classification can improve
disease control and epidemic response~\citep{ali2023benchmarking,chourasia2023efficient}.

\begin{figure}[h!]
    \centering
    \includegraphics[scale=0.30]{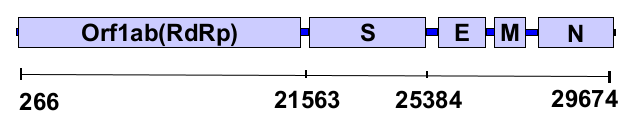}
    \caption{The coronavirus genome is 26--32kb in length. The
    structural genes include spike (S), envelope (E), membrane (M),
    and nucleocapsid (N). S region encodes the spike protein.}
    \label{fig_spike_seq_example}
\end{figure}

The coronaviruses (CoVs) are grouped into five genera, infecting
different hosts, including humans, palm civets, bats, dogs, and
monkeys, among others~\citep{li2006animal}. CoVs are known to mutate
quickly and adapt to new environments. They have shown a capacity for
animal-to-human, human-to-animal, and animal-to-animal
transmission~\citep{graham2010recombination}. There have been accounts
of cross-species transmission and alteration in viral tropism
resulting in new diseases in different
hosts~\citep{shi2008review,vijgen2006evolutionary}. The surface (S)
protein or spike protein of different CoVs is key to the binding and
entry of the virus into the host cell and determines the range of host
specificity. It is composed of the receptor-binding domain or S1
subunit and S2 subunit (see Figure~\ref{fig_spike_seq_example}) that
harbor sequences for viral fusion to the cell
membrane~\citep{li2006animal}. The spike proteins of CoVs recognize
different receptors across different hosts. Also, the sequences of the
S1 subunit of CoVs has been reported to show differences across
genera~\citep{li2016structure}.

\begin{figure}[h!]
    \centering
    \includegraphics[scale=0.30]{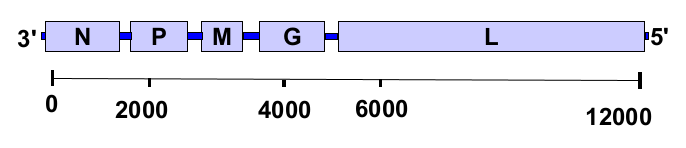}
    \caption{The rabies genome is 12kb in length and encodes five
    proteins Nucleoprotein (N), Phosphoprotein (P), Matrix Protein
    (M), Glycoprotein (G), and Polymerase (L).}
    \label{fig_rabies_genome}
\end{figure}

The rabies Lyssavirus (RABV), belongs to the genus Lyssavirus in
the Rhabdoviridae family.  Rhabdoviruses are simple viruses that
encode five proteins and appear as bullet-shaped, enveloped virions
with glycoprotein spikes on the surface.
These virions have a helical nucleocapsid within the envelope that is
symmetrically coiled into a cylindrical structure. The nucleocapsid is
composed of one molecule of negative sense, single-stranded RNA about
12kb long. Even with only five proteins encoded, the virus can protect
itself from ribonuclease digestion and retain a shape ideal for
transcription. Five proteins (N, P, M, G, and L) are produced as shown
in Figure~\ref{fig_rabies_genome}.

Traditional methods based on phylogenetic tree construction are
computationally expensive and do not scale to the large volume of sequence data~\citep{hadfield2018a}. Employing machine learning on sequencing data is a viable
alternative~\citep{ali2021spike2vec}.  However, some of the existing
sequence classification methods, such as the one proposed
in~\citep{kuzmin2020machine}, require the sequences to be aligned
(sequence characters must be in a one-to-one correspondence). Aligning
large volumes of sequences is computationally expensive (if required)
and utilizes expert knowledge that can potentially introduce bias in
the data~\citep{golubchik2007mind}.  
Planning future endemic/pandemic protection measures on time may be aided by
alignment-free embedding techniques~\citep{ali2023viralvectors}. They will be helpful in swiftly
implementing machine learning solutions and work as excellent tools
for healthcare professionals.

This paper proposes a feature vector generation named Virus2Vec. We depict the spike protein sequences of the SARS-CoV-2 and rabies viral nucleotide sequence data using Virus2Vec. It provides for improved host identification and downstream clustering and classification activities. Virus2Vec combines the use of
\emph{minimizers} and the position weight matrix (PWM) for a compact
alignment-free representation of amino acid sequences.  
Although the notion of minimizer is previously used in
metagenomics~\citep{girotto2016metaprob, chourasia2023reads2vec}, it has not been used (to the
best of our knowledge) for viral sequence classification. The main
contributions of this work are:
\begin{enumerate}
\item We propose Virus2Vec, a compact alignment-free embedding approach based on minimizers and the position weight matrix to
  generate a feature vector representation of different coronaviruses
  and rabies virus sequence data.
\item Our method eliminates the need for the sequence alignment step
  (multiple sequence alignment is an \textsc{NP-Hard}
  problem~\citep{chatzou2016multiple}) from the classification pipeline
  (unlike~\citep{kuzmin2020machine,ali2022pwm2vec}) while maintaining
  the performance of the underlying classifiers.
\item We show that without aligning the sequences and using a fraction
  of the information as compared to a more traditional $k$-mers based
  approach~\citep{ali2023benchmarking}, we are still able to outperform the
  baselines and state-of-the-art (SOTA) methods.
\item Virus2Vec is a compact sequence representation scheme that is
  scalable to ``Big Data'' and can also be used for many
  other sequence analysis tasks.
\end{enumerate}

Our manuscript is organized as follows: Section~\ref{sec_related_work}
contains the previous work for sequence
classification. Section~\ref{sec_proposed_approach} contains the
detail about our proposed alignment-free method for sequence
classification. Section~\ref{sec_experimental_setup} contains the
experimental setup and dataset collection and statistics detail. The
results for our proposed method are in
Section~\ref{sec_results_discussion}. Finally, we conclude our paper
in Section~\ref{sec_conclusion}.

\section{Related Work}~\label{sec_related_work}
After the spread of COVID-19, efforts have been made to study the virus's behavior using machine-learning approaches to biological sequences. 
Using a trait-based approach, authors in~\citep{worsley2020using} identified candidate wildlife species that may contribute to the transmission and maintenance of rabies Lyssavirus(RABV). This approach has a problem since the domestic dog (Canis lupus familiaris) is regarded as a key reservoir in many developing nations, particularly in African and Asian countries~\citep{cleaveland2017rabies}. Because of the vast number of canine cases and the absence of standard wildlife monitoring systems~\citep{vercauteren2012rabies} or diagnostic assays, most wildlife species' contributions to the preservation of certain RABV variants are still largely unknown~\citep{cordeiro2016importance}. Considering modern approaches, some work is done using sequence data for machine learning-based solutions.

Authors in~\citep{ali2023viralvectors} use $k$-mers and a kernel-based approach to classifying the spike sequences. However, it can not scale on big data because of memory inefficiency.
Authors in~\citep{kuzmin2020machine} propose using one-hot encoding to classify the viral hosts of coronavirus using spike sequences only. Although they achieved higher predictive performance, authors in~\citep{ali2023pssm2vec} show that the $k$-mers-based approach outperforms the one-hot. Authors in~\citep{taslim2023hashing2vec,ali2023anderson} propose a faster method for embedding generation but it mainly focuses on faster implementation as compared to generating a compact and effective embedding.

For the embedding generation of short reads data, authors in~\citep{chourasia2023reads2vec} advise using a minimizer-based technique. Additionally, the classification of metagenomic data has been suggested in~\citep{wood-2014-kraken,kawulok-2015-cometa}. To obtain accurate read binning for metagenomic data, the authors in~\citep{girotto2016metaprob} use probabilistic sequence signatures. According to the theoretical work on minimizers~\citep{universalHitting}, there is a close relationship between universal hitting sets and minimizers schemes, where efficient (low-density) minimizers schemes match up with small-sized universal hitting sets.  The main issue with all of these methods is that is intended for short reads data and they cannot be used in real-world situations where there may be millions of sequences because they cannot scale to larger datasets.

Sequence analysis, motif predictions, and identification investigations have effectively used position weight matrix (PWM) based techniques. A number of well-known software programs or web servers, such as the PWMscan software package \citep{ambrosini2018pwmscan}, and PSI-BLAST \citep{bhagwat2007psi}, have been developed based on the implementation of PWMs. The development of a PWM-based approach for protein function prediction and a justification for the PWM and its related characteristics' high potential for protein sequence analysis are presented in~\citep{cheol2010position}. Although the aforementioned approaches are effective in these fields, they do not offer a universal approach for designing a feature embedding for the underlying sequence, which would contain rich information about the sequence and serve as input to various machine learning algorithms.

In another work a position weight matrix (PWM) based approach is proposed in~\citep{ali2022pwm2vec}, which generates a fixed-length representation of spike sequences based on weights of $k$-mers computed using a PWM. However, their method only works with aligned sequence data.
Authors in~\citep{girotto2016metaprob} propose the use of minimizers for metagenomic data.
Since metagenomic data contains short reads, each can be represented by a single minimizer ($m$-mer)~\citep{chourasia2022clustering,chourasia2023reads2vec}.
Their approach is not directly applicable to our scenario.

\section{Proposed Approach}\label{sec_proposed_approach}
This section discusses our proposed alignment-free methods based on minimizers and the position weight matrix (PWM) to design a better feature vector representation from spike amino acid sequences and rabies virus nucleotide sequences. The problem of sequence classification is challenging due to the following points.
\begin{enumerate}
    \item Sequences can have different lengths. Designing a fixed-length numerical representation without loss of information becomes challenging.
    \item Mutations (changes in the sequence) do not happen randomly but rather due to selection pressures.  For example, mutations happen disproportionately many in the spike region of coronaviruses due to their importance in interfacing with the host. Designing a model to capture those variations is challenging.
    \item Some of the existing method requires sequence alignment as a preprocessing step. Designing a scalable alignment-free method without compromising on predictive performance is challenging.
\end{enumerate}

\subsection{Virus2Vec}
Although $k$-mers-based frequency vectors are proven to be efficient and perform better than the traditional one-hot-encoding on aligned sequences~\citep{ali2023viralvectors}, a major problem with
$k$-mers is that there are too many (similar) $k$-mers generated for a given sequence~\citep{wood2014kraken}.  Counting these similar $k$-mers can be an expensive --- and redundant --- task, as for each $k$-mer, we need to check which ``bin'' of the frequency vector it will be
placed. Another issue could be 
storing all $k$-mers in memory, especially for longer sequences.  Hence, we need memory efficient way to make the overall algorithm scalable. 
For a given $k$-mer, a \emph{minimizer} of length $m$ ($m < k$) is the $m$-mer that is lexicographically smallest both in forward and reverse order of the $k$-mer.  
\begin{remark}
Authors in~\citep{singh2017gakco} considered the first $m$ characters from $k$-mers (to design $m-mers$) rather than selecting the lexicographically smallest $m$ characters. However, we noted that we were getting better results by considering the smallest $m$ characters lexicographically. Thus, we use this approach.
\end{remark}

\begin{figure}
    \centering
    \includegraphics[scale = 0.16] {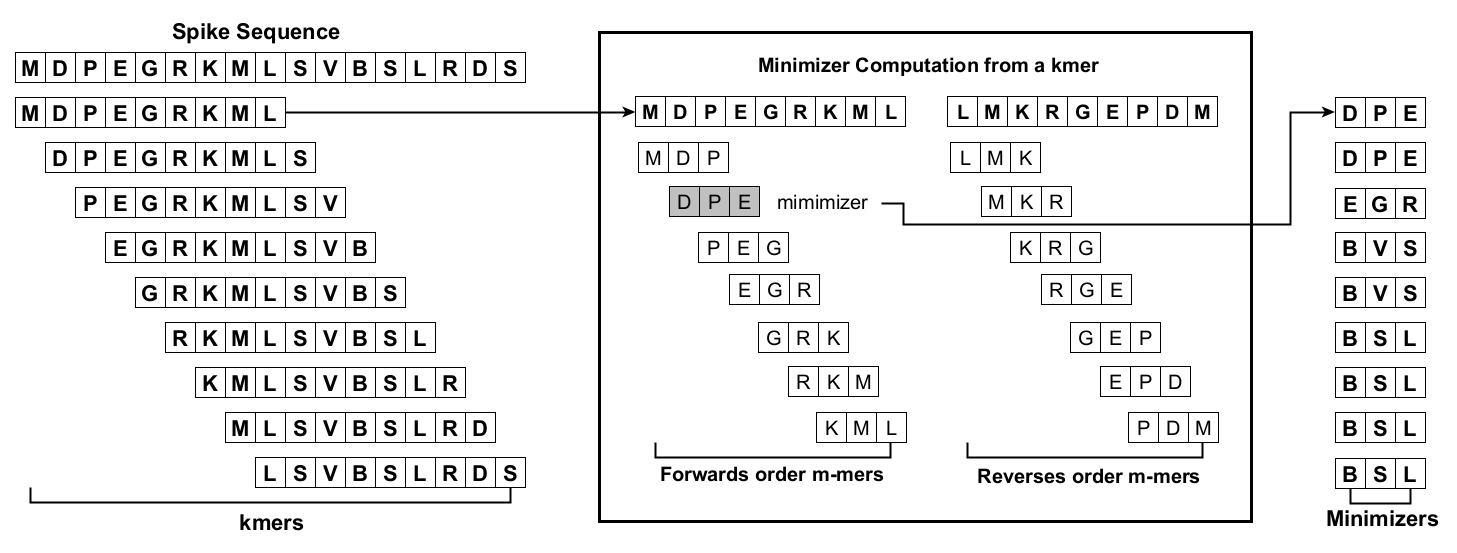}
    \caption{Example of $k$-mers and their corresponding minimizers in a spike amino acid sequence ``MDPEGRKMLSVBSLRDS". 
    }
    \label{fig_k_mers_minimizer}
\end{figure}

\begin{figure}[h!]
    \centering
    \includegraphics[scale=0.28]{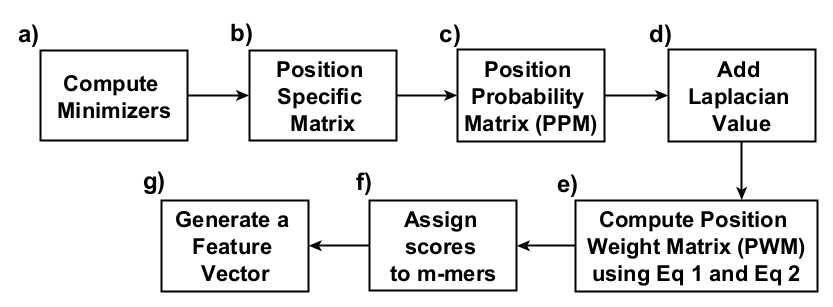}
    \caption{Virus2Vec Flow Diagram. 
    }
    \label{fig_Virus2Vec_flow_diagram}
\end{figure}

See Figure~\ref{fig_k_mers_minimizer} for an example of a set of $k$-mers and the corresponding minimizers. To compute minimizer, sliding window is used on $k$-mer to find minimizes from both directions(forward and reverse). And finally, the lexicographic smallest is selected as the minimizer for that $k$-mer.In this way, minimizers ignore many amino acids in each $k$-mer, only preserving a fraction of the $m$-mers, for which binning of these $m$-mers becomes much more efficient.  
Using the minimizers for  $m = 3$, which is decided empirically using standard validation set approach~\citep{validationSetApproach}, we generate a fixed-length feature vector of length $\vert \Sigma \vert^m$. 
For each minimizer, we compute a weight using the ``Position Weight Matrix" (PWM) method. Figure~\ref{fig_Virus2Vec_flow_diagram} shows the flow diagram for Virus2Vec.

Figure~\ref{fig_Virus2Vec_flow_diagram}, consists of the steps (a--g) as explained next. Given an input spike protein sequence, Figure~\ref{fig_Virus2Vec_flow_diagram} (a) extract the minimizers ($m$-mers) of length $3$ (decided using a standard validation set approach~\citep{validationSetApproach}).
A Position Frequency Matrix (PFM) is generated see in Figure~\ref{fig_Virus2Vec_flow_diagram} (b), which contains the frequency count for each character at each position.  
In our experiments, since we have $20$ unique amino acids in the spike protein sequence dataset, our PFMs have 20 rows and $m=3$ columns. Whereas for rabies virus sequence dataset 
we have $4$ unique nucleotide; our PFMs have 4 rows, and $m=3$ columns. In Figure~\ref{fig_Virus2Vec_flow_diagram} (c), we normalize the PFM matrix to create a Position Probability Matrix (PPM) containing the probability of each amino acid at each position.

It is possible that the frequency (hence probability in the PPM) of a character at a certain position is 0. To avoid zeros, we add a Laplace estimator or a pseudocount to each value in the position probability matrix as shown in Figure~\ref{fig_Virus2Vec_flow_diagram} (d). We use a pseudocount of 0.1 in our experiments~\citep{nishida2009pseudocounts}. A position weight matrix (PWM)is then computed from the adjusted probability matrix (after adding laplacian). We make the PWM by computing the log-likelihood of each amino acid character $c$, i.e., $c \in A, C,\dots,Y$ for spike sequences or  $c \in A, C,G,T$ for rabies virus sequences, appearing at each position $i$ according to the following expression:
\begin{equation}
\label{eq_pwm}
    W_{c, i}=\log_{2} \frac{p(c, i)}{p(c)} 
\end{equation}
where $c \in A, C...Y(bases)$ or $c \in A, C, G, T(bases)$ and 
\begin{equation}
p(c) = \frac{n(c)}{61}
\end{equation}
The $n(c)$ is the number of codons for each amino acid (i.e., 1 for MW, 2 for CFYHQNKDE, 3 for I, 4 for VPTAG, and 6 for RL) and $61$ is the number of sense codons.

As shown in Figure~\ref{fig_Virus2Vec_flow_diagram} (e), by using Equation~\ref{eq_pwm}, a scalar value to each Amino Acid (AA) for each position in $m$-mer is assigned.
After getting the PWM, we use it to compute the absolute scores for each individual minimizer generated from the sequence.  
It is the sum of the score of bases for the index. 

After getting the score for each $m$-mer, the final step as shown in Figure~\ref{fig_Virus2Vec_flow_diagram} (g) we generate a vector of length $\vert \Sigma \vert^m$. 
We use the score of each $m$-mer (computed using the PWM-based approach) to the corresponding bin to get the final feature vector representation.
The pseudocode for Virus2Vec is given in Algorithm~\ref{algo_peptide}. 

\begin{remark}
Note that steps b to f in Figure~\ref{fig_Virus2Vec_flow_diagram} for our method are the same as given in PWM2Vec~\citep{ali2022pwm2vec}. Our method differs in a way that it works with the minimizers instead of $k$-mers (our input is different as given in step a). The idea of using minimizers is that they are proven to work better compared to the $k$-mers in the metagenomic domain~\citep{girotto2016metaprob}. However, their use for full-length sequences is not well explored. 
Similarly, our feature embedding is ``general" and can work for both aligned and unaligned sequences (see step g), unlike the PWM2Vec, which only works with aligned sequences. Our methods differ in the computation of likelihood weight for each amino acid, where we consider $log_2 \frac{p(c, i)}{p(c)}$ rather than the equal probability of each amino acid (which is $\frac{1}{Unique\_AA\_Count}$) as given in PWM2Vec method.
\end{remark}

\begin{algorithm}[h!]
    \caption{Virus2Vec Overall Computation}
    \label{algo_peptide}
    \begin{algorithmic}
        \State \textbf{Input:} Set of Spike Sequences $S$ of dimension $X \times Y$, alphabet $\Sigma$, $k$-mer length $k$, m-mer length $m$
	\State \textbf{Output:} Weighted Frequency Matrix $V$
        \Function{CompFreqVector}{$S, \Sigma, k$}
        \State $V$ = [] \Comment{Weighted Freq Matrix}   
        
        \For{i $\leftarrow 1$ to $ \vert X \vert$}{
            \State $\mathbb{A}$ = \Call{CompMinimizer}{$s,k,m$}
            \Comment{$ \vert \Sigma \vert \times k$ Matrix }
            \State $PFM$ = 0 * [$ \vert \Sigma \vert$] [$k$]
			
			\For{p $\leftarrow 0$ to $ k$}
                {
				\State  $PFM[:,p]$ = \Call{GetAlphabetCount}{$\mathbb{A}$[:,p]}
                }
			
			\State PC =$0.1$ \Comment{Pseudocount}
			\State PPM = \Call{CompProbability}{PFM} + PC
			\State p(c) = n(c) / 61  
			\State PWM = $\frac{PPM}{p(c)}$ 
			\State W = []
   
			\For{u $\leftarrow 0$ to $ \vert \mathbb{A} \vert$}{
				\State  W.append(\Call{CompMmersScore}{$\mathbb{A}$[u]})
    }
			\State combos = \Call{GenAllCombinations}{$\Sigma, k$}
			\State $v$ = [0] * $\vert \Sigma \vert^{k}$ 
   
            \For{i $\leftarrow 1$ to $ \vert \mathbb{A} \vert$}{
                \State idx = combos.index($\mathbb{A}$[i]) 
				\Comment{\text{find $i^{th} kmer$ idx}}
                \State $v$[idx] $\leftarrow$ $V$[idx] + W[i]
                }
			\State V.append($v$)
   }
    \State \Return $V$
    \EndFunction
	\end{algorithmic}
\end{algorithm}

\section{Experimental Setup}\label{sec_experimental_setup}
This section describes the setup we used for the experiments, followed by the dataset statistics in the ~\ref{Sec_Data_Stats}. We also give a visual representation of the data using t-SNE plots and its discussion in the ~\ref{Sec_Data_visualization}. Later, we discuss the baseline models in ~\ref{Sec_Baseline_models} and its detailed discussion in ~\ref{sec_baseline_description} followed by we discussion of the Ablation study in Section~\ref{Sec_Ablation_Study}.

All experiments are conducted using an Intel(R) Xeon(R) CPU E7-4850 v4
@ $2.10$GHz having Ubuntu $64$ bit OS ($16.04.7$ LTS Xenial Xerus)
with 3023 GB memory.  
The algorithm is implemented in Python, and the code is available online for
reproducibility~\footnote{\url{https://github.com/sarwanpasha/Virus2Vec}}.
We use several algorithms and metrics for classification, as shown in Table~\ref{tbl_no_feat_select_results_unaligned}.
We are using the area under the curve (AUC) receiver operating characteristic curve AUC-ROC because it is prone to overestimation 
and pulls the false positive rate towards zero~\citep{sofaer2019area}. Finally, we report each classifier's training time (in seconds). We split the data into $70-30\%$ training and testing (held-out) sets, respectively.
We run experiments with $5$ random initialization for train-test splits and report average results. We use $5$ fold cross-validation in the training set for hyperparameter tuning.

\subsection{Dataset Collection and Statistics}
\label{Sec_Data_Stats}
In this work, we use two datasets: Spike Sequence from the SARS-CoV-2 virus, and Rabies sequences data. The statistics and details are provided in Table~\ref{tbl_data_summary}. The multiple sequence alignment (MSA) is conducted using Mafft Alignment software to get aligned sequences for Coronavirus Host Data that we use to evaluate the baseline approach, some of which require Aligned Sequences.

\begin{table*}[h!]
  \centering
  \resizebox{0.6\textwidth}{!}{
  \begin{tabular}{L{1.7cm}L{2.5cm}L{2.2cm}p{1cm}C{1cm}C{0.9cm}C{0.9cm}cC{0.9cm}} 
    \toprule
    \multirow{2}{*}{Name} & \multirow{2}{*}{Type} & \multirow{2}{*}{Source} & \multirow{2}{1.4cm}{Sequence Count} & \multirow{2}{*}{Classes} & \multicolumn{4}{c}{Sequence Length} \\
    \cmidrule{6-9}
    & & & & & Min & Max & Avg & Mode \\
    \midrule	\midrule	
    Coronavirus Host Data & Spike protein sequences for COVID-19 hosts  & GISAID~\citep{gisaid_website_url}, ViPR~\citep{pickett2012vipr} & 5558 & 22 & 9 & 1584 & 1272.4 & 1273 \\
    \cmidrule{2-9}
    Rabies Virus Data & Nucleotide genome sequences for rabies virus hosts & RABV-GLUE~\citep{RABV_GLUE_website_url} & 20051 & 12 & 90 & 11930 & 1948.4 & 1353\\
    \bottomrule
  \end{tabular}
  }
  \caption{Data Statistics.
  }
  \label{tbl_data_summary}
\end{table*}

\subsection{Data Visualization}
\label{Sec_Data_visualization}
\begin{figure*}[!ht]
\floatconts
  {fig:subfigex}
  {\caption{t-SNE plots for \textbf{Coronavirus Host data} ($5558$ sequences) different feature embeddings.The figure is best seen in color.}
  \label{host_tsne}}
  {%
    \subfigure[Spike2Vec]{\label{}%
      \includegraphics[width=0.20\linewidth]{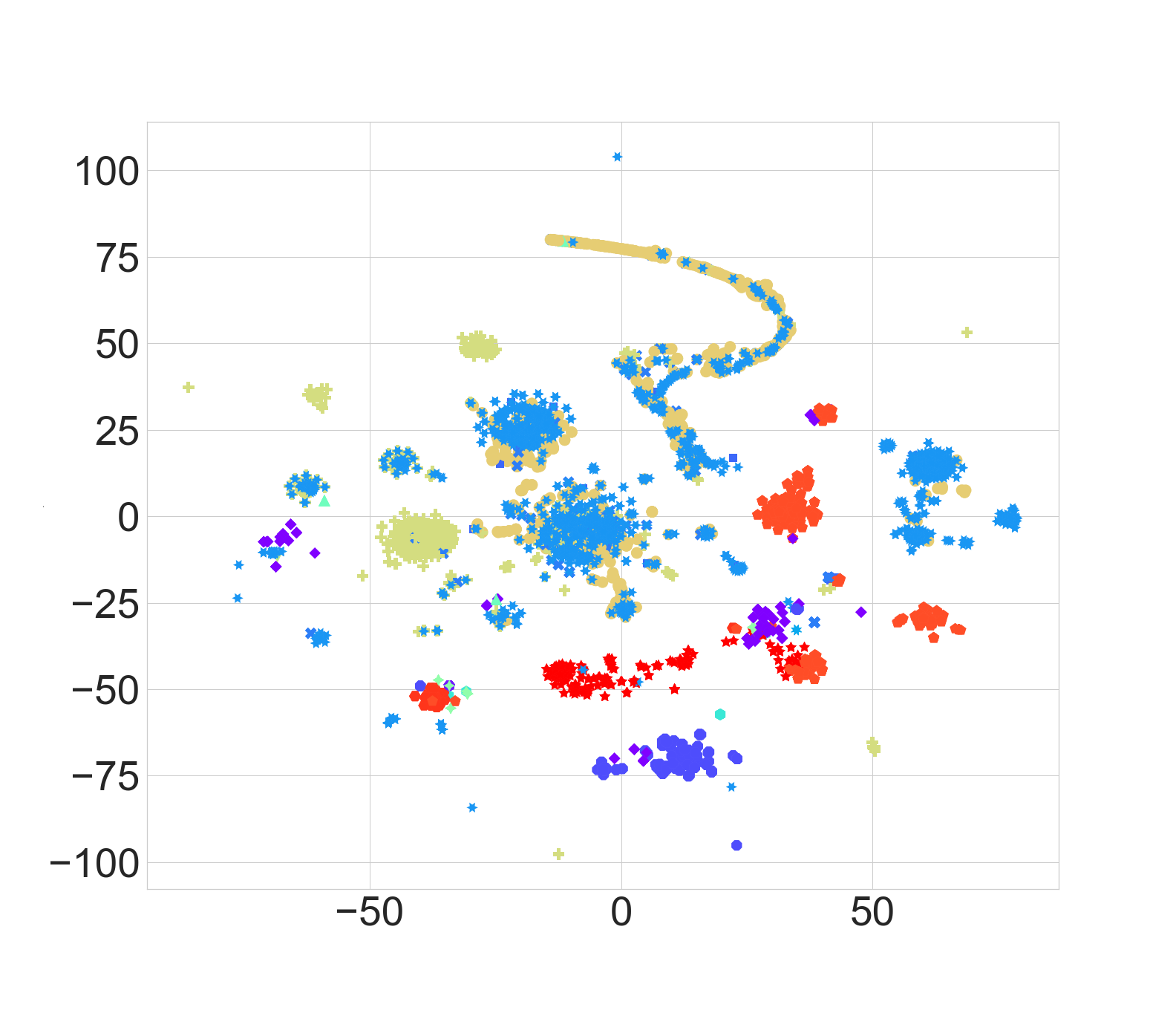}}%
    \subfigure[Approx. Kernel]{\label{}%
      \includegraphics[width=0.20\linewidth]{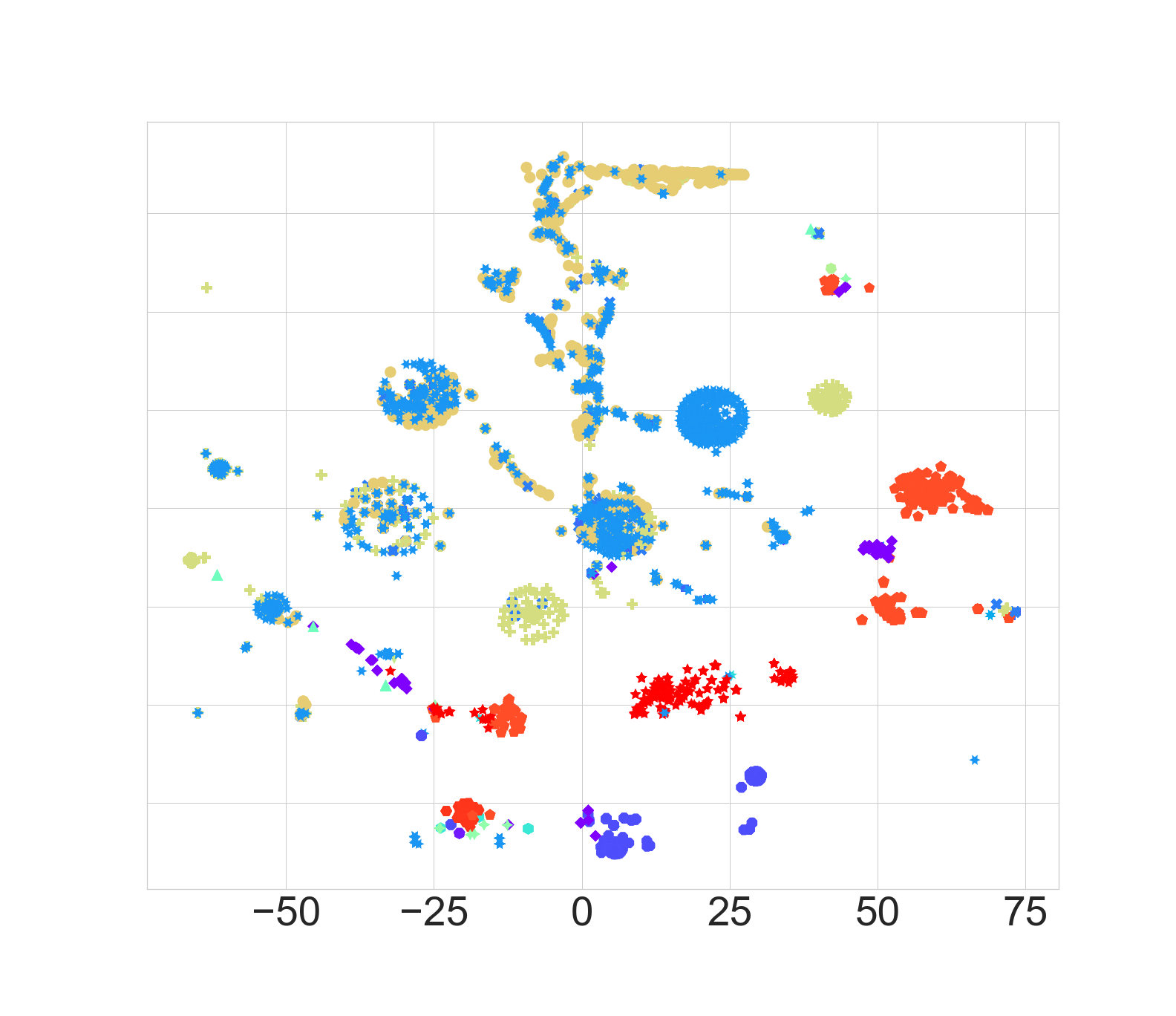}}%
    \subfigure[MFV]{\label{}%
      \includegraphics[width=0.20\linewidth]{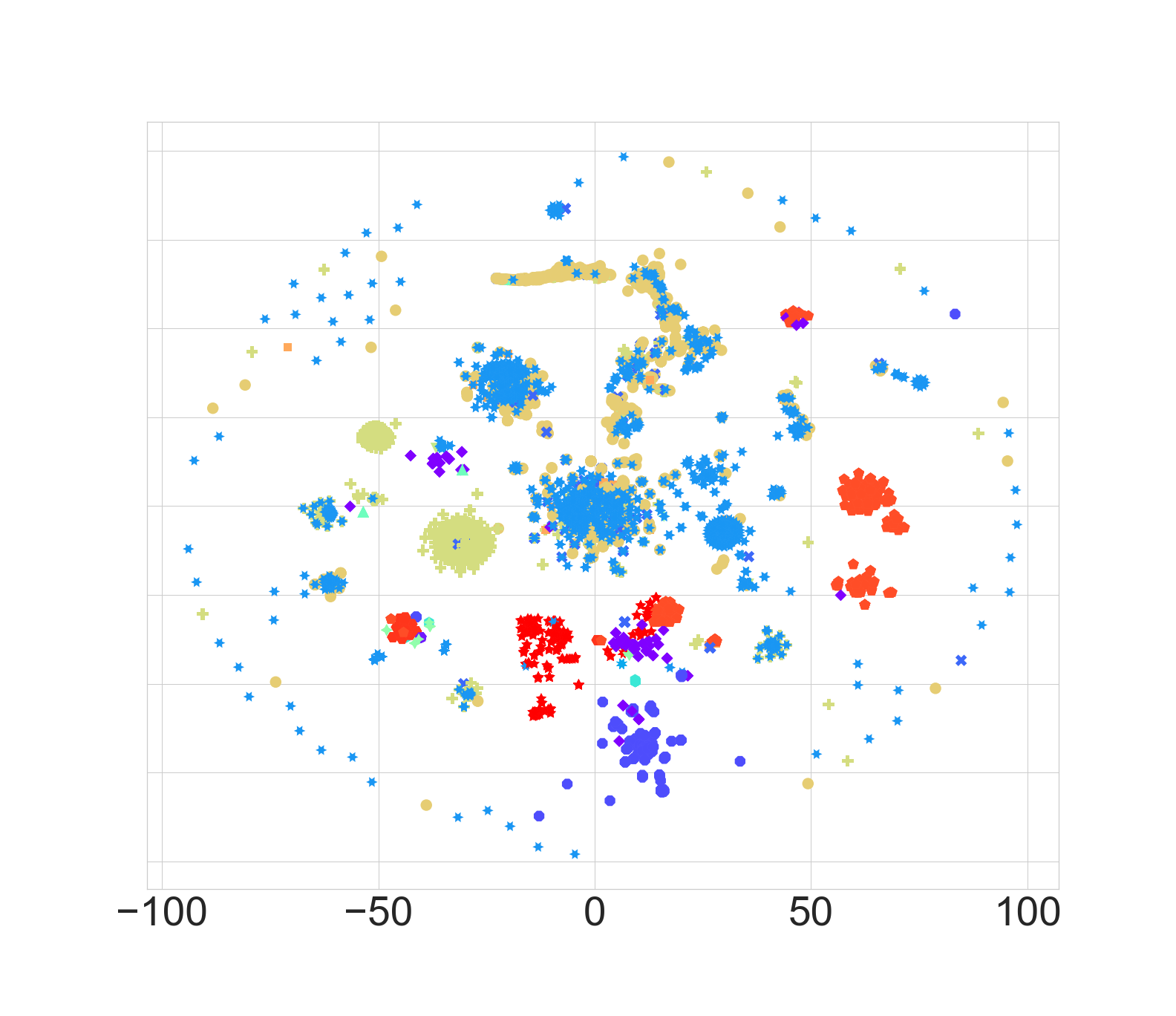}}%
    \subfigure[PWM2Vec]{\label{}%
      \includegraphics[width=0.20\linewidth]{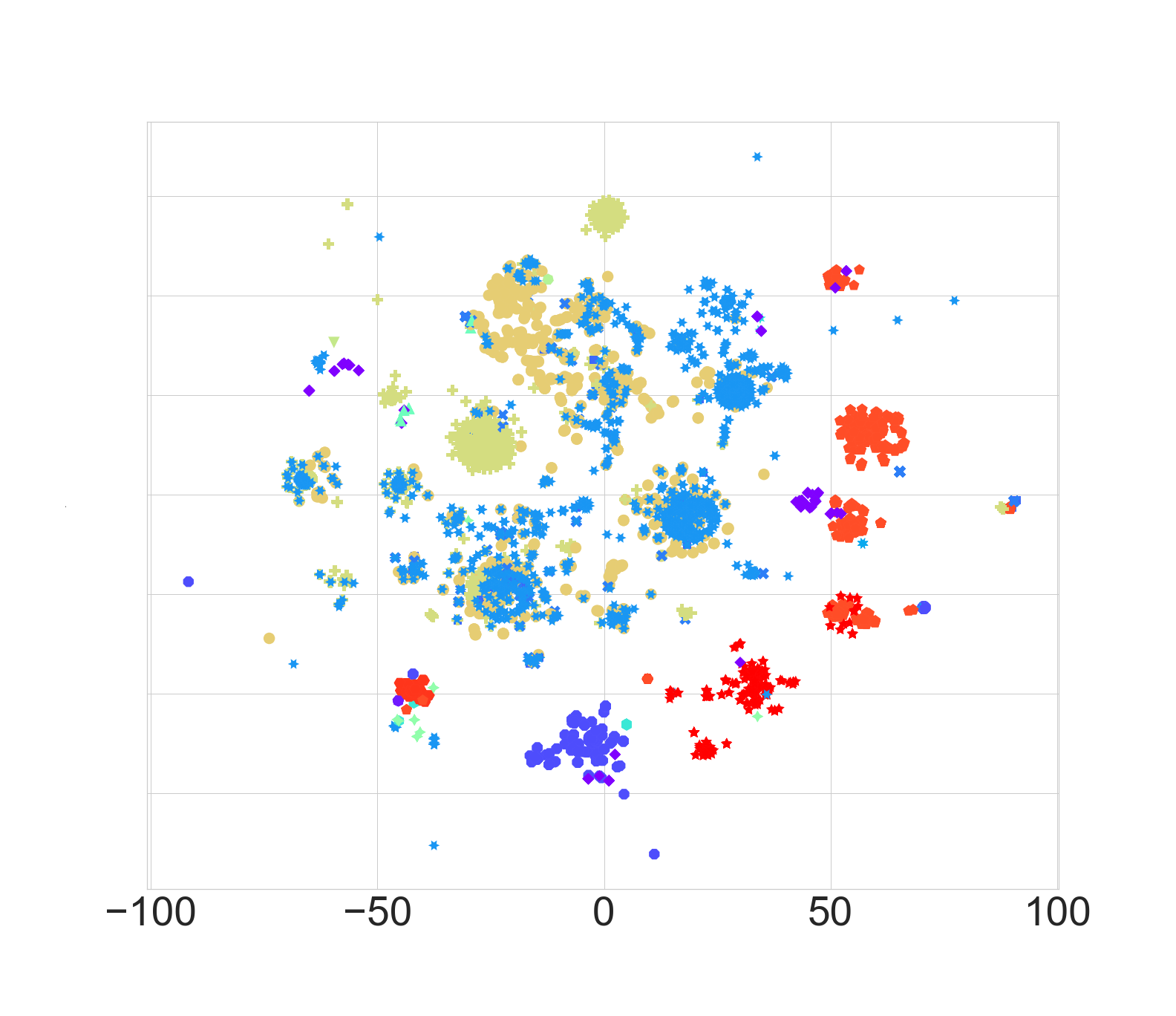}}%
    \subfigure[Virus2Vec]{\label{}%
      \includegraphics[width=0.20\linewidth]{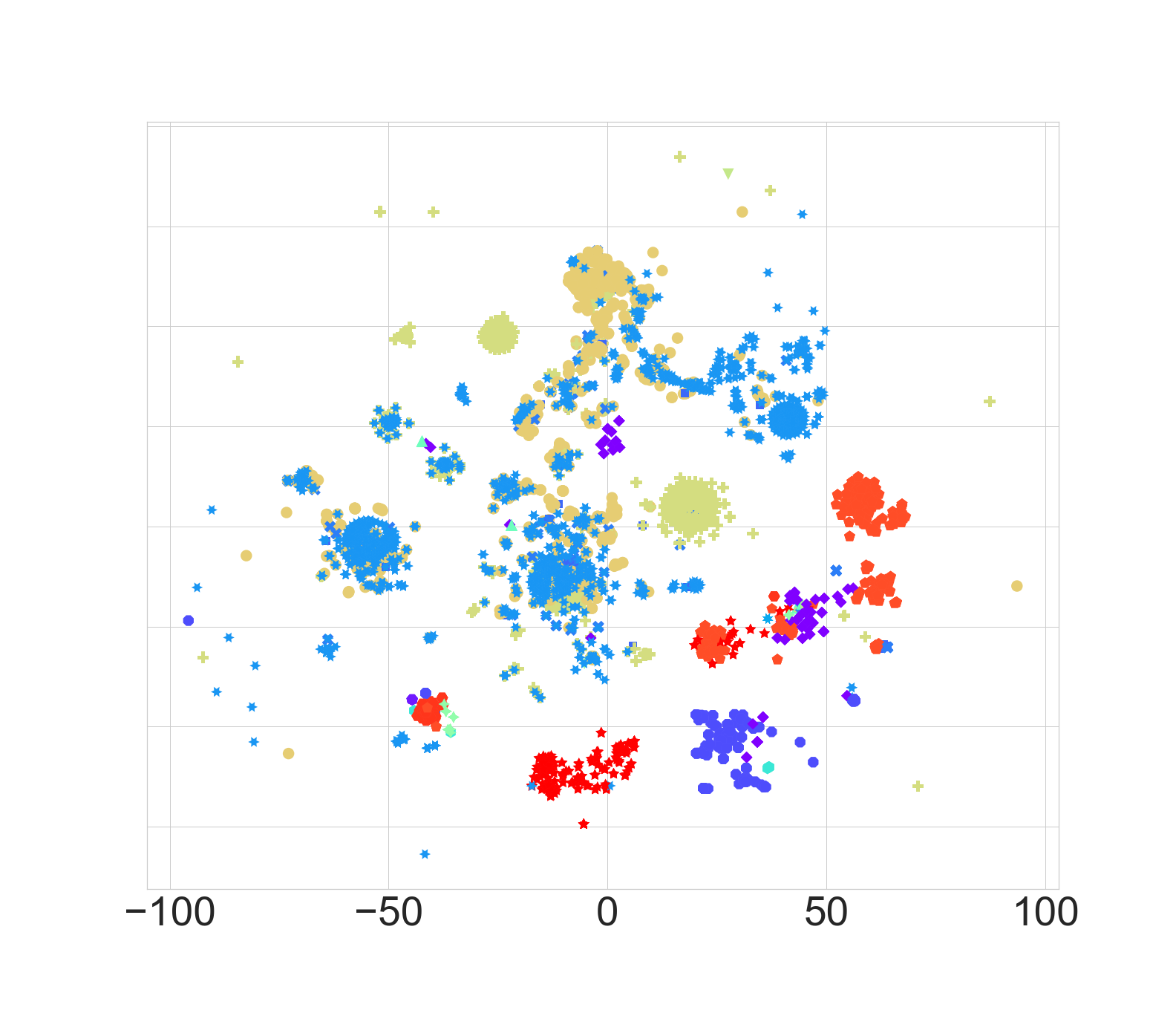}}%

      \includegraphics[width=0.8\linewidth]{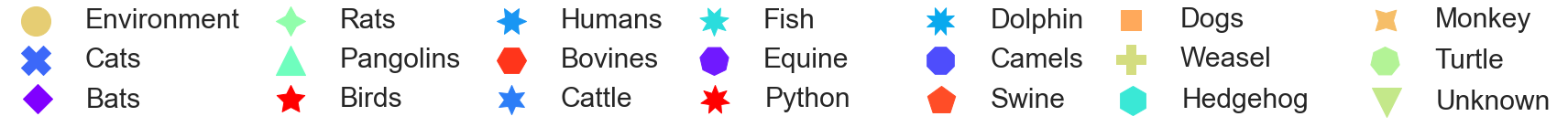}}%
\end{figure*}

\begin{figure*}[!ht]
\floatconts
  {fig:subfigex}
  {\caption{t-SNE plots for \textbf{Rabies Virus} ($20051$ sequences) for different feature embeddings.The figure is best seen in color.}
  \label{host_tsne_rabies}}
  {%
    \subfigure[Spike2Vec]{\label{Spike2Vec_rabies}%
      \includegraphics[width=0.20\linewidth]{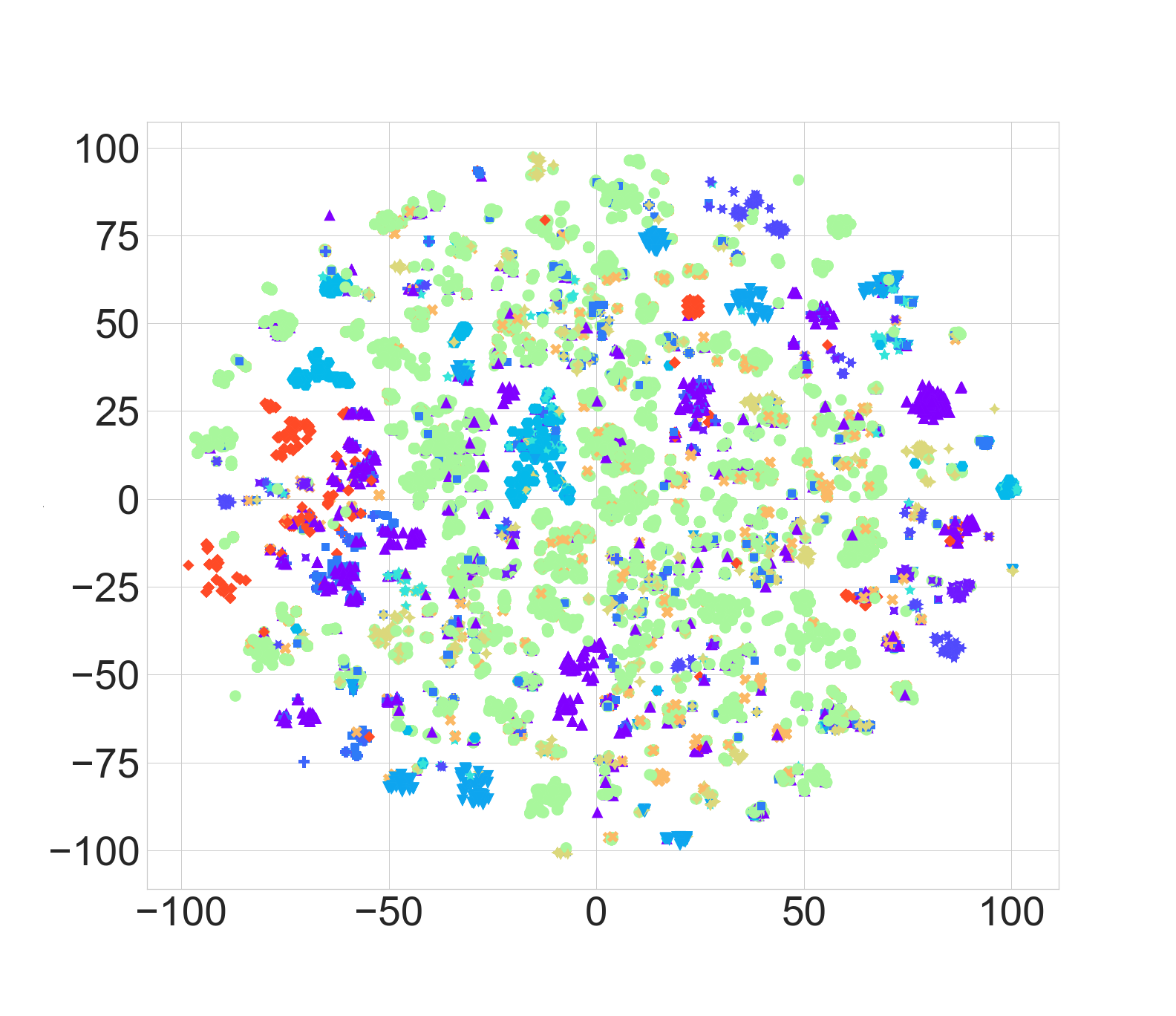}}%
    \subfigure[Approx. Kernel]{\label{Approx_rabies}%
      \includegraphics[width=0.20\linewidth]{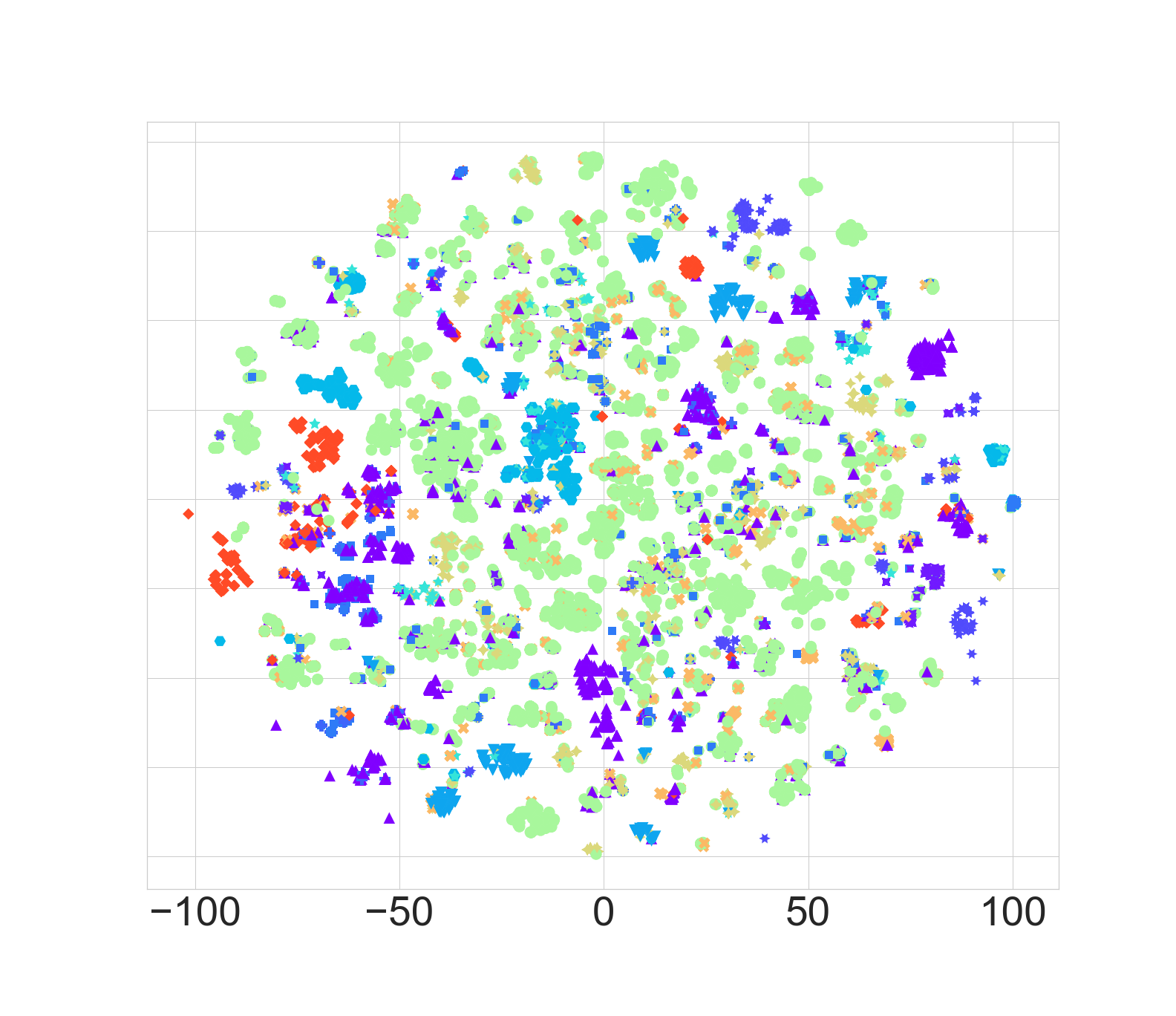}}%
    \subfigure[MFV]{\label{Min2Vec_rabies}%
      \includegraphics[width=0.20\linewidth]{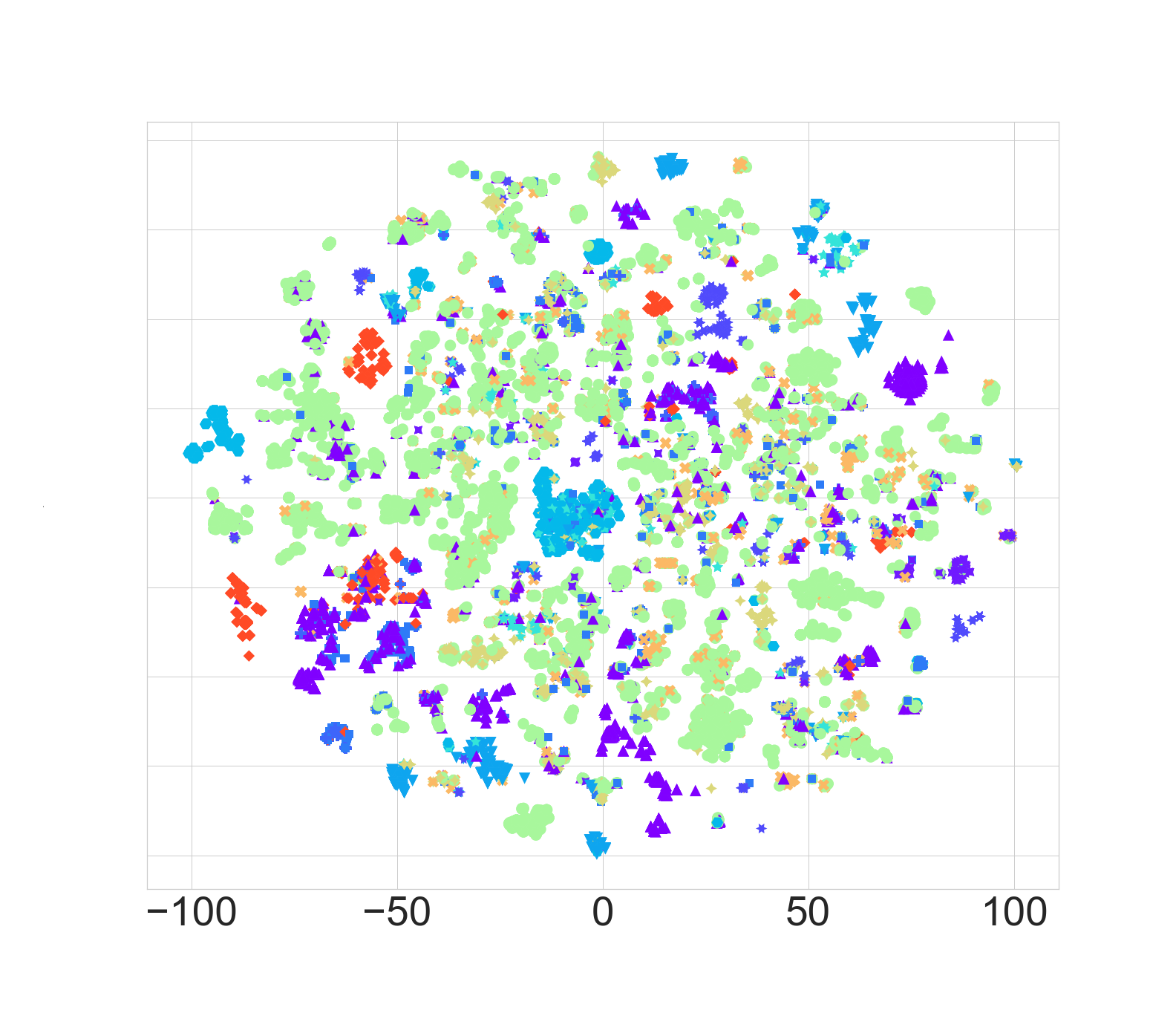}}%
    \subfigure[PWM2Vec]{\label{PSWM2Vec_rabies}%
      \includegraphics[width=0.20\linewidth]{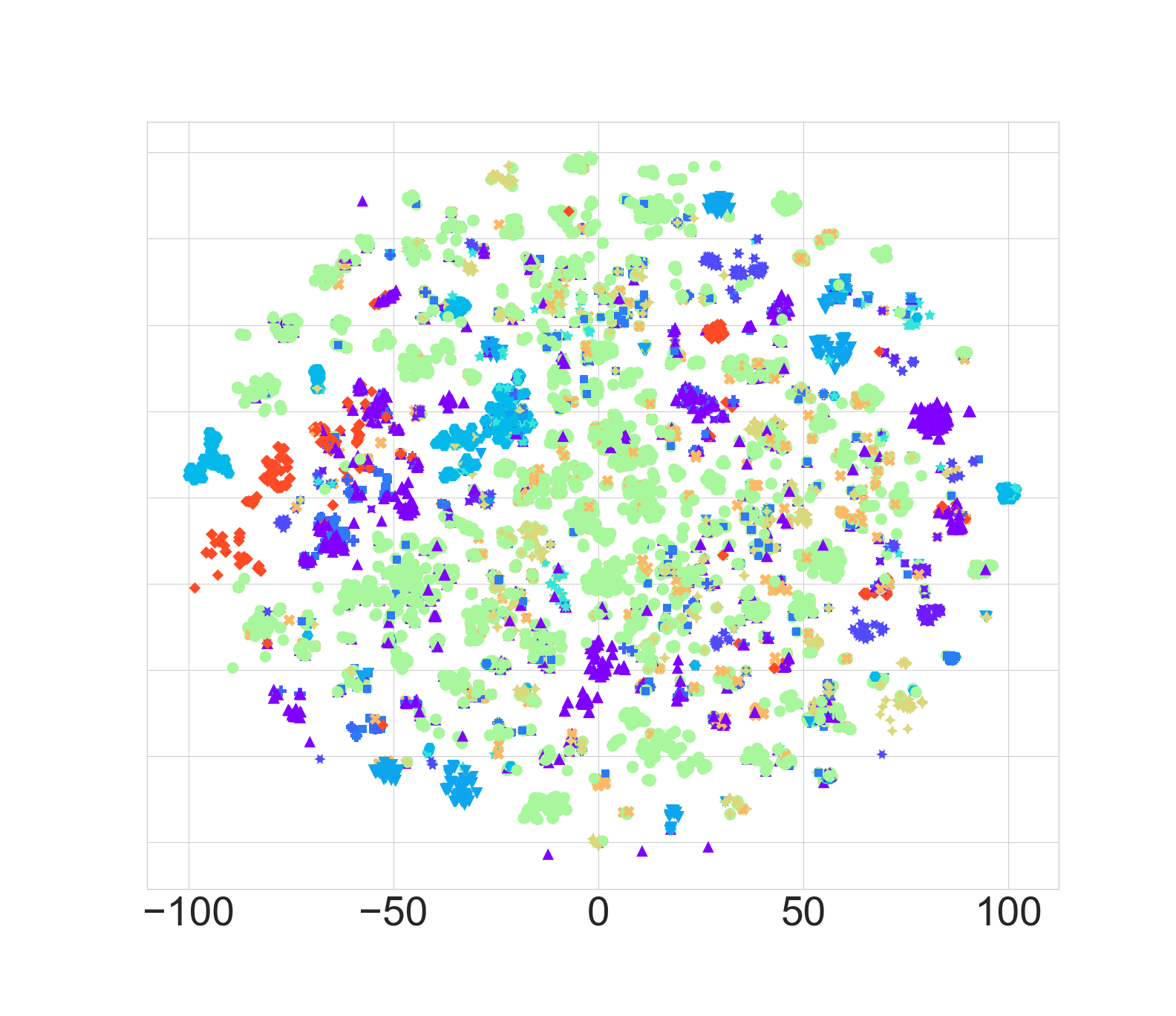}}%
    \subfigure[Virus2Vec]{\label{Peptide2Vec_rabies}%
      \includegraphics[width=0.20\linewidth]{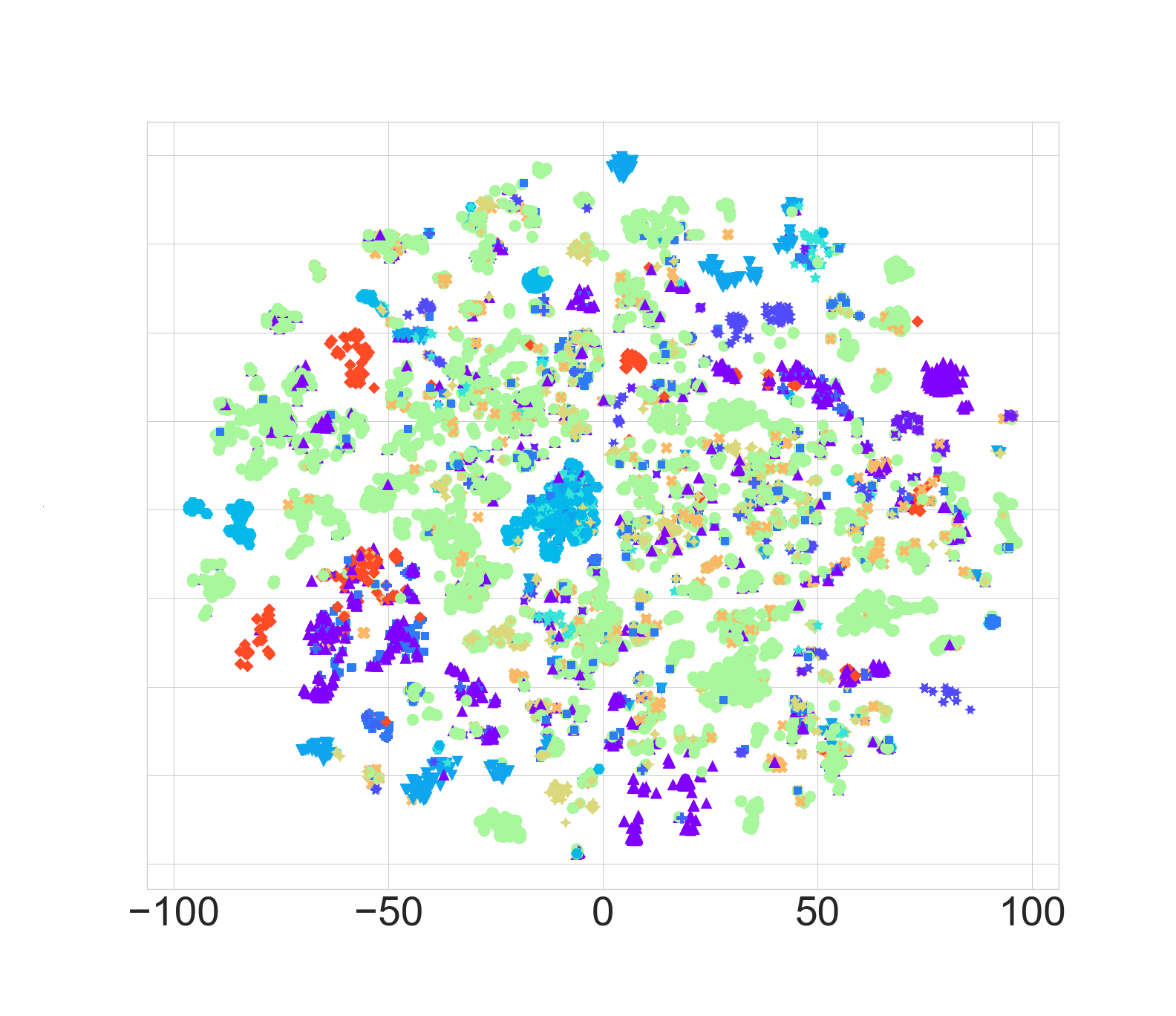}}%
    
      \includegraphics[width=0.80\linewidth]{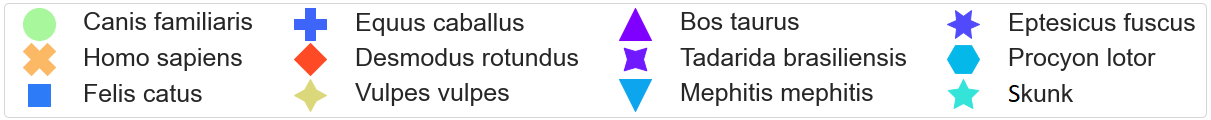}}%
\end{figure*}

In order to see if there is any natural (hidden) clustering in the data, we use t-distributed stochastic neighbor embedding (t-SNE)~\citep{van2008visualizing}, which maps input sequences to 2D representation. Particularly in the natural sciences, t-SNE is well-liked because of its ability to handle vast volumes of data and its usage for dimensionality reduction while maintaining the structure of the data~\citep{chourasia2022informative}. Here, we use t-SNE to analyze and contrast the ability of various embeddings to preserve the structure. The t-SNE plots for SARS-CoV-2 spike sequences for different embedding methods are shown in Figure~\ref{host_tsne} for Spike2Vec, Approx. Kernel, MFV, PWM2Vec, and Virus2Vec, respectively. We can observe that with Virus2Vec, t-SNE is able to preserve the structure of data in the same way as with the other existing embedding methods, which shows that Virus2Vec is preserving the overall structure of data. Similarly, Figure~\ref{host_tsne_rabies} shows the t-SNE plots for rabies virus data for mentioned embeddings. Similar results can be seen in spike data. Virus2Vec does not disturb the structure and even provides better clusters as compared to baseline embeddings.

\subsection{Baseline Models}
\label{Sec_Baseline_models}
A brief description of baselines and their comparison with the proposed model are provided in table~\ref{tbl_baseline_andproposed_summary}. Detail description is given in section~\ref{sec_baseline_description}. We are using $3$ Neural Network models the configuration is provided in Table~\ref{tbl_baseline_NN_summary}.

\begin{table*}[h!]
  \centering
  \resizebox{0.6\textwidth}{!}{
  \begin{tabular}{L{2cm}C{1.5cm}L{3cm}L{1.8cm}L{2cm}C{1.4cm}C{2cm}C{1.2cm}} 
    \toprule
Model & Embedding Layers & Custom Model Layer & Activation Function & Middle Layer & Dense Layer Dimension & Output Layer & Optimizer \\
    \midrule	\midrule	
Long Short-Term Memory (LSTM)~\citep{hochreiter1997long} & 500 & 2 LSTM layer 200 Units & Leaky Layer with alpha=0.05 & dropout with 0.2 & 500 & output layer Sigmoid Activation Function & ADAM Optimizer \\
\cmidrule{2-8}
Gated Recurrent Unit (GRU)~\citep{cho2014properties} &  500 & A GRU layer 200 Units & Leaky Layer with alpha=0.05 & dropout with 0.2 & 500 & Sigmoid Activation Function & ADAM Optimizer \\
\cmidrule{2-8}
Convolutional Neural Network (CNN)~\citep{lee2017human} & 500 & $2 \times 1-D$ convolutional layer(Conv1D) with 128 Filters and Kernel size 5 & Leaky Layer with alpha=0.05 & Batch Normalization, a max pooling layer of size 2 & 500 & Sigmoid Activation Function & ADAM Optimizer \\ 
    \bottomrule
  \end{tabular}
  }
  \caption{Baseline neural network configuration.}
  \label{tbl_baseline_NN_summary}
\end{table*}

\subsection{Baseline Methods Description}
\label{sec_baseline_description}

\paragraph{One-Hot Encoding (OHE)}
A fixed-length numerical feature vector, called OHE is proposed in~\citep{kuzmin2020machine}. It generates a binary (0-1) vector based on the character's position in the sequence given $\Sigma$. Since the length of each spike sequence (after alignment) in our data is $3498$, the length of OHE for a spike sequence is then $3498 \times 20 = 69,960$.

\paragraph{Spike2Vec}
The spike sequence classification method, Spike2Vec, is recently proposed in~\citep{ali2021spike2vec}. Given a sequence, Spike2Vec computes $N$ $k$-mers, where $N= L - k + 1$ ($L$ is the length of the spike sequence and $k=3$ as given in~\citep{ali2021spike2vec}). 
After generating the $k$-mer for a spike sequence, the count of each $k$-mer is used to get the frequency vector.

\paragraph{Approximate Kernel}
A kernel-based method for sequence classification is proposed in~\citep{ali2022efficient} (for reference, we call this method ``Approx. Kernel"). It computes the  kernel value between two sequences using the dot product based on the matches and mismatches among the $k$-mers spectrum. This kernel matrix is then used in our case as input to kernel PCA~\citep{hoffmann2007kernel} to acquire the feature vector representation.

\paragraph{PWM2Vec} 
PWM2Vec~\citep{ali2022pwm2vec} assigns different weights to each $k$-mer in the feature vector depending on the values of the characters in the position weight matrix. The feature vector length equals the total number of $k$-mers in an aligned sequence. 
The length of PWM2Vec-based embedding equals the number of $k$-mers in a sequence. Since the number of $k$-mers could be different for different length (unaligned) sequences, this method only applies to aligned sequences. Since the length of sequences in our dataset is $3498$ (after alignment), the feature vector length for PWM2Vec is $3490$ (which is equal to the total number of $k$-mers in a sequence, where $k=9$ as mentioned in~\citep{ali2022pwm2vec}).

\paragraph{Spaced $k$-mers}
The feature embeddings generated by $k$-mers suffer from sparsity and the curse of dimensionality, which negatively impacts analytical performance. To address these issues, the concept of spaced $k$-mers was introduced~\citep{singh2017gakco}. Spaced $k$-mers are a set of non-contiguous substrings of length $k$, also known as $g$-mers. To generate an embedding using spaced $k$-mers, the $g$-mers of the sequence are first computed, and then $k$-mers are extracted from these $g$-mers. In our experiments, we use $k=4$ and $g=9$, which were selected using a standard validation set approach.

We transform the raw sequence data into the numerical form that the machines can process to build the neural network (NN) models (for end-to-end training on the sequences). We use one hot encoding (with data padding) method considering 20 amino acids in the data.  
We use the following NN models as described below.
 
\subsubsection{Long Short-Term Memory (LSTM)}
The architecture of \textsc{lstm} scheme~\citep{hochreiter1997long} consists of an embedding layer (of embedding is $500$), an LSTM layer with $200$ memory units, a LeakyReLU layer with alpha = $0.05$, an LSTM layer again with $200$ memory units followed by another LeakyReLU layer, a dropout with value $0.2$, a Dense layer of dimensions $500$ followed by LeakyReLU layer, and finally an output layer and a sigmoid activation function. We use the ADAM~\citep{adam} optimizer.

\subsubsection{Gated Recurrent Unit (GRU)}
The GRU model~\citep{cho2014properties} consists of an embedding layer (of embedding is $500$), a GRU layer with $200$ memory units, a LeakyReLU layer with alpha = $0.05$ followed by a Dropout layer with value $0.2$, and finally a dense output layer and a sigmoid activation function. We also use the ADAM optimizer in the GRU architecture.

\subsubsection{Convolutional Neural Network (CNN)}
The CNN model~\citep{lee2017human} comprises an embedding layer (of embedding is $500$), a 1-D convolution layer (Conv1D) with $128$ filters with a kernel size of $5$, a LeakyReLU layer with alpha = $0.05$, a batch normalization layer, a 1-D convolution layer (Conv1D) again with $128$ filters and a kernel size of $5$, a LeakyReLU layer with alpha = $0.05$ followed by batch normalization, a max pooling layer (pool size of $2$), a dense layer of 500 dimensions followed by a LeakyReLU layer with alpha = $0.05$, and finally an output dense layer with a sigmoid activation function. For optimization, we use the ADAM optimizer.

\begin{table*}[h!]
  \centering
  \resizebox{0.7\textwidth}{!}{
  \begin{tabular}{L{3cm}C{2cm}L{2cm}C{2cm}C{1.6cm}C{2cm}L{4.5cm}} 
    \toprule
    Embedding & Alignment Free & Low Dim Vectors & $\vert$ Vector $\vert$ (Spike/Rabies) & Space Efficient & Runtime Efficient & Details\\
    \midrule	\midrule	
    One-Hot Encoding~\citep{kuzmin2020machine} & \xmark & \xmark & 69960 / 5600 & \xmark & \xmark & length of OHE for a spike sequence $3498$ \\
    \cmidrule{2-7}
    Spike2Vec~\citep{ali2021spike2vec} & \cmark & \cmark & 8000 / 125 & \cmark & \xmark & $\Sigma = 20$ and $k=3$ (for Spike data) \\
     \cmidrule{2-7}
    Approx. Kernel~\citep{farhan2017efficient} & \cmark & \cmark & 500 / 500 & \xmark & \xmark & Dimensionality depends on Num of sequences \\
     \cmidrule{2-7}
    PWM2Vec~\citep{ali2022pwm2vec} & \xmark & \cmark & 3490 / 125 & \cmark & \cmark & Length of Spike Seq after alignment $3498$ and $k=9$ \\
     \cmidrule{2-7}
    LSTM~\citep{hochreiter1997long} & \cmark & \_ & \_ & \xmark & \xmark & \multirow{6}{4cm}{End-to-End deep learning architectures that take one-hot encoding representation as input and perform classification} \\
     \cmidrule{2-6}
    GRU~\citep{cho2014properties} & \cmark & \_ & \_ & \xmark & \xmark &  \\
     \cmidrule{2-6}
    CNN~\citep{lee2017human} & \cmark & \_ & \_ & \xmark & \xmark &  \\
     \cmidrule{2-7}
    ProteinBert~\citep{brandes2022proteinbert} & \cmark & \_ & \_ & \xmark & \xmark & Pretrained Protein language model using Transformer  \\
     \cmidrule{2-7}
    MFV & \cmark & \cmark & 8000 / 125 & \cmark & \cmark & $k=9$ and $m=3$ \\
     \cmidrule{2-7}
   \cmidrule{2-7}
    Virus2Vec (ours) & \cmark & \cmark & 8000 / 125 & \cmark & \cmark & Proposed method $m=3$ \\    
    \bottomrule
  \end{tabular}
  }
  \caption{Baseline and Proposed Methods advantages and disadvantages. 
  }
  \label{tbl_baseline_andproposed_summary}
\end{table*}

\subsection{Ablation Study}
\label{Sec_Ablation_Study}
\paragraph{MFV} Minimizer-based feature vector is a method in which for a given $k$-mer, a \emph{minimizer} of length
$m$ ($m < k$) is computed, called as $m$-mer which is  a lexicographically smallest in forward and reverse order of the $k$-mer (See Figure~\ref{fig_k_mers_minimizer} for an example).  
Fixed-length frequency vector from the set of minimizers, which contains each minimizer's count in the frequency vector's corresponding bin. The pseudocode for generating the frequency vectors is given in Algorithm~\ref{algo_freq_vec}. Given $\Sigma$ and length of minimizers $m$ (where $m = 3$), the length of each vector is  $\vert \Sigma \vert^m$.

\begin{algorithm}[h!]
    \caption{Build Frequency vector from $k$-mers and/or Minimizers}
    \label{algo_freq_vec}
    \textbf{Input:} Set $\mathcal{S}$ of kmers($k$-mer) or Minimizers (m-mers) on alphabet $\Sigma$, the size of mers $n$ (k for kmer or m for minimizer) \Comment{$\text{n represents $k$ or $m$}$}\\
    \textbf{Output:} Frequency Vector $V$   

    \begin{algorithmic}
    \Function{GetFrequencyVector}{$\mathcal{S},n,\Sigma$}
	\State combos = \Call{GenAllPossibleNMers}{$\Sigma$, n}     
        \State $V$ = [0] * $\vert \Sigma \vert^{n}$ 
	\Comment{$\text{zero vector}$} 
        \State \For{ i $\leftarrow 1$ to $ \vert \mathcal{S} \vert$}{\
            \State idx = combos.index($\mathcal{S}$[i]) \Comment{Find $i^{th}  m-mer$ idx}
            \State $V$[idx] $\leftarrow$ $V$[idx] + 1 
            \Comment{$\text{Increment bin by 1} $}
        } 
        \State \Return $V$
    \EndFunction
    \end{algorithmic}
\end{algorithm}

\paragraph{PSWM2Vec}
A method based on Position Specific Weight Matrix (PSWM) is similar to PWM2Vec with two differences. First, use minimizers instead of the $k$-mers to initiate the PWM (as given in Figure~\ref{fig_Virus2Vec_flow_diagram}). Secondly, after generating PWM, we flatten the matrix to get a feature vector whose length will equal the number of entries in PWM.
The pseudocode for PSWM2Vec is given in Algorithm~\ref{algo_pswm2vec}. 
The results for MFV, PSWM2Vec, and Virus2Vec for aligned data are given in Table~\ref{tbl_no_feat_select_results_aligned} and for unaligned in Table~\ref{tbl_no_feat_select_results_unaligned}. Virus2Vec outperforms the other two embeddings for all but one evaluation metric. 

\begin{algorithm}[h!]
    \caption{PSWM2Vec algorithm pseudocode}
    \label{algo_pswm2vec}
    \textbf{Input} Set of Spike Sequences $S$, alphabet $\Sigma$, $k$-mer length $k$, m-mer length $m$\\
    \textbf{Output} Weighted Frequency Matrix $V$
    
    \begin{algorithmic}
    \Function{ComputeFreqVector}{$S, \Sigma, k$} 
        \State $V$ = [] \Comment{Weighted Frequency Matrix}
        
        \State \For{i $\leftarrow 1$ to $ \vert S.rows \vert$}{ 
            \State mList = \Call{CompMinimizer}{$s,k,m$}
            \State $PFM$ = $0$ * [$ \vert \Sigma \vert$] [$k$]
			
			\State \For{p $\leftarrow 0$ to $ k$}{
				\State  $PFM[:,p]$ = \Call{GetAlphbtCnt}{mList[:,p]}
			}
			
			\State PC =$0.1$ \Comment{Pseudocount}
			\State PPM = \Call{CompProbability}{PFM} + PC
			\State $v$ = flat(PPM)
			\Comment{Flat PWM to get vector $v$}
			\State V.append($v$)
        }
    \State \Return $V$
    \EndFunction
	\end{algorithmic}
\end{algorithm}


        
			
			

\section{Results and Discussion}\label{sec_results_discussion}
In this section, we present results for proposed embeddings and compare their performance with the baseline and SOTA methods (with and without sequence alignment).   

Our results show that not only Virus2Vec outperforms the SOTA methods but also preserves structure when compared with subjective evaluation using t-SNE plots. The runtime to generate the embeddings makes it a huge factor in considering Virus2Vec over other embeddings. 

Table~\ref{tbl_no_feat_select_results_unaligned}, which shows the results for different embedding methods (on unaligned data) with different classification algorithms. We can observe that the Virus2Vec outperforms the SOTA methods in terms of different evaluation metrics. Note that since OHE and PWM2Vec work with aligned data only, we have not included those for unaligned data results. The performance of Virus2Vec is not very different for aligned and unaligned data. Other methods, such as Approx. the kernel has better ROC-AUC values for aligned data as compared to unaligned ones. Spike2Vec, on the other hand, was able to generalize better on unaligned data. We can observe that Virus2Vec outperforms not only the feature
engineering-based baselines but also the neural network-based
classifiers. The NN models are known for their application in image data. Also, since the data is of a smaller size it is likely that
the NN models failed to learn the patterns in the sequences. The findings are reckoned by its visualization counterpart as well, as we saw in t-SNE plots also for Virus2Vec, it does not disrupt the general structure of the data because t-SNE is able to retain the structure of the data in the same way that the other embedding methods do.

\begin{table*}[h!]
    \centering
    \resizebox{0.96\textwidth}{!}{
    \begin{tabular}{@{\extracolsep{6pt}}cp{1.1cm}p{0.9cm}p{0.9cm}p{1.1cm}p{1.2cm}p{1.2cm}p{1.1cm}p{1.4cm}|
    p{0.9cm}p{0.9cm}p{1.1cm}p{1.2cm}p{1.2cm}p{1.1cm}p{1.4cm}}
    \toprule
    & & \multicolumn{7}{c}{Host Spike Sequences} & \multicolumn{7}{c}{Rabies Virus} \\
    \cmidrule{3-9} \cmidrule{10-16}
        \multirow{2}{*}{Method} & \multirow{2}{*}{Classifier}  & \multirow{2}{*}{Acc. $\uparrow$} & \multirow{2}{*}{Prec. $\uparrow$} & \multirow{2}{*}{Recall $\uparrow$} & \multirow{2}{1.2cm}{F1 (Weig.) $\uparrow$} & \multirow{2}{1.4cm}{F1 (Macro) $\uparrow$} & \multirow{2}{1.2cm}{ROC AUC $\uparrow$} & Train Time (sec.) $\downarrow$
          & \multirow{2}{*}{Acc. $\uparrow$} & \multirow{2}{*}{Prec. $\uparrow$} & \multirow{2}{*}{Recall $\uparrow$} & \multirow{2}{1.2cm}{F1 (Weig.) $\uparrow$} & \multirow{2}{1.4cm}{F1 (Macro) $\uparrow$} & \multirow{2}{1.2cm}{ROC AUC $\uparrow$} & Train Time (sec.) $\downarrow$\\
        \midrule \midrule
        \multirow{7}{1.8cm}{Spike2Vec}
          & SVM & 0.84 & 0.84 & 0.84 & 0.83 & 0.77 & 0.87 & 45.36 & 0.72 & 0.70 & 0.72 & 0.69 & 0.58 & 0.76 & 22.76 \\
        & NB & 0.69 & 0.77 & 0.69 & 0.67 & 0.58 & 0.79 & 6.02 & 0.06 & 0.29 & 0.06 & 0.03 & 0.03 & 0.52 & 0.40   \\
        & MLP & 0.81 & 0.83 & 0.81 & 0.81 & 0.63 & 0.83 & 46.14 & 0.58 & 0.45 & 0.58 & 0.48 & 0.23 & 0.60 & 1.46  \\
        & KNN & 0.80 & 0.81 & 0.80 & 0.79 & 0.59 & 0.79 & 1.97 & 0.75 & 0.73 & 0.75 & 0.74 & 0.62 & 0.79 & 1.07  \\
        & RF & 0.84 & 0.85 & 0.84 & 0.84 & 0.73 & 0.85 & 10.21 & 0.78 & 0.76 & 0.78 & 0.76 & 0.67 & 0.81 & 0.88   \\
        & LR & 0.84 & 0.85 & 0.84 & 0.84 & 0.76 & 0.87 & 31.00 &  0.71 & 0.67 & 0.71 & 0.67 & 0.55 & 0.75 & 1.14   \\
        & DT & 0.82 & 0.83 & 0.82 & 0.82 & 0.71 & 0.85 & 2.54 & 0.68 & 0.68 & 0.68 & 0.68 & 0.57 & 0.77 & 0.21   \\
        \midrule
        \multirow{7}{1.8cm}{Approx. Kernel}
          & SVM & 0.79 & 0.80 & 0.79 & 0.77 & 0.57 & 0.78 & 18.18  & 0.73 & 0.72 & 0.73 & 0.71 & 0.59 & 0.76 & 244.82	\\
        & NB & 0.60 & 0.66 & 0.60 & 0.57 & 0.51 & 0.73 & \textbf{0.07} & 0.14 & 0.51 & 0.14 & 0.13 & 0.20 & 0.60 & 0.33       \\
        & MLP & 0.79 & 0.78 & 0.79 & 0.78 & 0.59 & 0.75 & 7.69 & 0.77 & 0.77 & 0.77 & 0.76 & 0.63 & 0.79 & 119.56    \\
        & KNN & 0.86 & 0.85 & 0.86 & 0.86 & 0.60 & 0.76 & 0.21 & 0.83 & 0.82 & 0.83 & 0.82 & 0.69 & 0.83 & 5.57      \\
        & RF & 0.82 & 0.82 & 0.82 & 0.81 & 0.67 & 0.78 & 1.80 & 0.83 & 0.83 & 0.83 & 0.82 & 0.71 & 0.83 & 22.17      \\
        & LR & 0.76 & 0.77 & 0.76 & 0.74 & 0.64 & 0.76 & 2.36 &  0.66 & 0.64 & 0.66 & 0.64 & 0.55 & 0.73 & 80.32      \\
        & DT & 0.78 & 0.78 & 0.78 & 0.77 & 0.55 & 0.75 & 0.24 & 0.76 & 0.76 & 0.76 & 0.76 & 0.65 & 0.80 & 4.44       \\
        \midrule
        \multirow{3}{2cm}{Neural Network}
        & LSTM & 0.32 & 0.10 & 0.32 & 0.15 & 0.02 & 0.50 & 21634.34 & 0.49 & 0.38 & 0.49 & 0.36 & 0.15 & 0.49 & 35026.49 \\
         & CNN 
        & 0.44 & 0.10 & 0.11 & 0.08 & 0.07 & 0.53 & 17856.40 & 0.73 & 0.74 & 0.73 & 0.72 & 0.64 & 0.80 & 8164.93 \\
         & GRU 
        & 0.32 & 0.13 & 0.32 & 0.16 & 0.03 & 0.50 & 126585.0 & 0.59 & 0.54 & 0.59 & 0.51 & 0.28 & 0.60 & 16180.78 \\
        \midrule
        \multirow{7}{1.6cm}{Spaced k-mer}
        & SVM & 0.81 & 0.82 & 0.81 & 0.81 & 0.89 & 0.92 & 3.12 & 0.80 & 0.80 & 0.80 & 0.79 & 0.67 & 0.81 & 140.67 \\
& NB & 0.66 & 0.69 & 0.66 & 0.66 & 0.61 & 0.78 & 0.03 & 0.28 & 0.56 & 0.28 & 0.27 & 0.35 & 0.69 & 0.11    \\
& MLP & 0.82 & 0.82 & 0.82 & 0.82 & 0.77 & 0.87 & 41.66 & 0.79 & 0.78 & 0.79 & 0.79 & 0.66 & 0.81 & 84.70  \\
& KNN & 0.79 & 0.80 & 0.79 & 0.80 & 0.77 & 0.87 & 0.40 & 0.83 & 0.82 & 0.83 & 0.82 & 0.71 & 0.84 & 2.54   \\
& RF & 0.84 & 0.85 & 0.84 & 0.84 & \textbf{0.91} & \textbf{0.94} & 2.84 & 0.84 & \textbf{0.84} & 0.84 & 0.83 & 0.72 & 0.84 & 24.28   \\
& LR & 0.82 & 0.83 & 0.82 & 0.82 & 0.89 & 0.93 & 2.31 & 0.79 & 0.78 & 0.79 & 0.78 & 0.66 & 0.81 & 14.08   \\
& DT & 0.80 & 0.80 & 0.80 & 0.80 & 0.85 & 0.93 & 0.64 & 0.76 & 0.76 & 0.76 & 0.76 & 0.65 & 0.81 & 6.36    \\
        \midrule 
        \multirow{1}{1.9cm}{Protein Bert}
 & \_ &  0.79 &  0.80 & 0.79 & 0.78 & 0.71 & 0.84 & 15742.95  &  0.79 & 0.78 & 0.79 & 0.76 & 0.64 & 0.80 & 35742.84 \\
 \cmidrule{1-9} \cmidrule{10-16}
        \multirow{7}{1.8cm}{MFV}
         & SVM & 0.83 & 0.83 & 0.83 & 0.82 & 0.73 & 0.85 & 35.71  & 0.66 & 0.61 & 0.66 & 0.61 & 0.48 & 0.71 & 241.11  \\
        & NB & 0.63 & 0.75 & 0.63 & 0.63 & 0.49 & 0.72 & 5.80  & 0.06 & 0.34 & 0.06 & 0.05 & 0.08 & 0.54 & 0.41     \\
        & MLP & 0.82 & 0.82 & 0.82 & 0.82 & 0.66 & 0.81 & 53.82  & 0.61 & 0.54 & 0.61 & 0.56 & 0.33 & 0.65 & 2.17    \\
        & KNN & 0.79 & 0.80 & 0.79 & 0.78 & 0.63 & 0.81 & 1.60  & 0.74 & 0.72 & 0.74 & 0.72 & 0.61 & 0.79 & 1.12    \\
        & RF & 0.84 & 0.85 & 0.84 & 0.84 & 0.74 & 0.85 & 10.79  & 0.78 & 0.77 & 0.78 & 0.76 & 0.66 & 0.80 & 0.81     \\
        & LR & 0.83 & 0.84 & 0.83 & 0.83 & 0.74 & 0.85 & 9.24  & 0.59 & 0.55 & 0.59 & 0.54 & 0.36 & 0.64 & 0.70     \\
        & DT & 0.83 & 0.83 & 0.83 & 0.82 & 0.74 & 0.85 & 1.15  & 0.69 & 0.68 & 0.69 & 0.69 & 0.58 & 0.77 & 0.19     \\
        \midrule
        \multirow{7}{1.8cm}{PSWM2Vec}
          & SVM & 0.81 & 0.82 & 0.81 & 0.80 & 0.80 & 0.90 & 3.46  & 0.48 & 0.28 & 0.48 & 0.33 & 0.08 & 0.48 & 1.10 \\
        & NB & 0.58 & 0.66 & 0.58 & 0.57 & 0.53 & 0.78 & 0.25  & 0.27 & 0.32 & 0.27 & 0.26 & 0.16 & 0.27 & 0.18 \\
        & MLP & 0.82 & 0.82 & 0.82 & 0.81 & 0.72 & 0.87 & 8.44  &  0.57 & 0.50 & 0.57 & 0.50 & 0.33 & 0.57 & 2.33 \\
        & KNN & 0.81 & 0.80 & 0.81 & 0.80 & 0.70 & 0.86 & 1.22  & 0.64 & 0.62 & 0.64 & 0.62 & 0.50 & 0.64 & 0.49 \\
        & RF & 0.85 & 0.85 & 0.85 & 0.84 & 0.83 & 0.91 & 1.26 & 0.66 & 0.65 & 0.66 & 0.65 & 0.53 & 0.66 & 0.79 \\
        & LR & 0.79 & 0.80 & 0.79 & 0.77 & 0.70 & 0.84 & 1.45  & 
        0.48 & 0.31 & 0.48 & 0.34 & 0.10 & 0.48 & 1.41 \\
        & DT & 0.80 & 0.81 & 0.80 & 0.80 & 0.73 & 0.88 & \textbf{0.23}  & 0.58 & 0.59 & 0.58 & 0.58 & 0.47 & 0.58 & 0.17 \\
        \midrule
        \multirow{7}{2cm}{Virus2Vec}
          & SVM & 0.85 & 0.86 & 0.85 & 0.85 & 0.87 & 0.932 & 151.5  & 0.66 & 0.62 & 0.66 & 0.62 & 0.50 & 0.72 & 15931.90	\\
        & NB & 0.67 & 0.78 & 0.67 & 0.65 & 0.65 & 0.83 & 5.67  & 0.07 & 0.34 & 0.07 & 0.05 & 0.10 & 0.55 & \textbf{0.17}         \\
        & MLP & 0.85 & 0.85 & 0.85 & 0.84 & 0.79 & 0.90 & 47.30  & 0.71 & 0.69 & 0.71 & 0.68 & 0.56 & 0.75 & 11.76       \\
        & KNN & 0.84 & 0.85 & 0.84 & 0.83 & 0.76 & 0.88 & 78.79  & 0.71 & 0.73 & 0.74 & 0.71 & 0.59 & 0.78 & 8.54    \\
        & RF & 0.86 & 0.86 & 0.86 & 0.85 & 0.84 & 0.91 & 13.36  & \textbf{0.84} & 0.83 & \textbf{0.84} & \textbf{0.83} & \textbf{0.74} & \textbf{0.85} & 3.13         \\
        & LR & \textbf{0.87} & \textbf{0.87} & \textbf{0.87} & \textbf{0.87} & 0.88 & 0.93 & 8.29  & 0.59 & 0.54 & 0.59 & 0.53 & 0.34 & 0.63 & 13.94        \\
        & DT & 0.81 & 0.82 & 0.81 & 0.81 & 0.76 & 0.88 & 2.49  & 0.77 & 0.77 & 0.77 & 0.77 & 0.68 & 0.82 & 0.55         \\
        \bottomrule
    \end{tabular}
    }
    \caption{Performance comparison for different embedding methods on \textbf{Unaligned Spike Sequence} and \textbf{Unaligned Rabies Virus} data. The best values are shown in bold.}
    \label{tbl_no_feat_select_results_unaligned}
\end{table*}

We show the runtime comparison for computing different feature embeddings in Table~\ref{tbl_embedding_runtime}. It can be seen in the results that Virus2Vec takes the least time to generate for $5558$ sequences as compared to other methods. It takes $4$ times less as compared to the Approximate Kernel method and 15 times less than PWM2Vec, which are comparable when accuracy is considered. 
Similarly for the Rabies virus-host dataset also embedding generation is very less in the case of Virus2Vec. Although One hot encoding is taking less time the pre-processing for sequence alignment is not considered here which will make it expensive.

\begin{table}[h!]
  \centering
  \resizebox{0.49\textwidth}{!}{
  \begin{tabular}{lcc}
    \toprule
    Method & Coronavirus data Runtime $\downarrow$ & Rabies virus data Runtime $\downarrow$ \\
    \midrule	\midrule	
    OHE & 196.31 Sec. & \textbf{44.17} Sec.\\
    Spike2Vec & 1179.66 Sec. & 259.86 Sec.\\
    PWM2Vec & 1506.63 Sec.  & 412.254 Sec.\\
    Approx. Kernel & 379.47 Sec. &  179.47 Sec.\\
    Virus2Vec & \textbf{90.65} Sec. & 105.78 Sec. \\
    \bottomrule
  \end{tabular}
  }
  \caption{Runtime for generating feature vectors using different embedding methods for Coronavirus-Host data and  Rabies Virus-Host dataset.
  }
  \label{tbl_embedding_runtime}
\end{table}

Table~\ref{tbl_no_feat_select_results_aligned} shows the results for different embedding (on aligned data) with different classifiers. Here also, Virus2Vec outperforms the feature engineering-based baselines and the neural network classifiers. We can see a significant improvement in accuracy when compared with the Approximate kernel and PWN2Vec. It is definitely an advantage but the reduced runtime for embedding generation, when compared with the closest comparable embeddings, is huge and makes it a better choice.

\begin{table}[h!]
    \centering
    \resizebox{0.5\textwidth}{!}{
    \begin{tabular}{cp{1.1cm}p{0.9cm}p{0.9cm}p{1.1cm}p{1.2cm}p{1.2cm}p{1.1cm}|p{1.4cm}}
    \toprule
        \multirow{2}{*}{DL Model} & \multirow{2}{*}{Method}  & \multirow{2}{*}{Acc. $\uparrow$} & \multirow{2}{*}{Prec. $\uparrow$} & \multirow{2}{*}{Recall $\uparrow$} & \multirow{2}{1.2cm}{F1 (Weig.) $\uparrow$} & \multirow{2}{1.4cm}{F1 (Macro) $\uparrow$} & \multirow{2}{1.2cm}{ROC AUC $\uparrow$} & Train Time (sec.) $\downarrow$ \\
        \midrule \midrule
        
        \multirow{7}{1.8cm}{OHE~\citep{kuzmin2020machine}}
        & SVM & 0.83 & 0.84 & 0.83 & 0.83 & 0.70 & 0.84 & 389.128 \\
        & NB & 0.67 & 0.80 & 0.67 & 0.65 & 0.64 & 0.82 & 56.741 \\
        & MLP & 0.77 & 0.76 & 0.77 & 0.75 & 0.44 & 0.71 & 390.289 \\
        & KNN & 0.80 & 0.79 & 0.80 & 0.79 & 0.55 & 0.78 & 16.211 \\
        & RF & 0.84 & 0.84 & 0.84 & 0.83 & 0.66 & 0.84 & 151.911 \\
        & LR & 0.84 & 0.85 & 0.84 & 0.83 & 0.71 & 0.85 & 48.786 \\
        & DT & 0.83 & 0.84 & 0.83 & 0.82 & 0.64 & 0.82 & 21.581 \\
        \midrule
        \multirow{7}{1.8cm}{Spike2Vec~\citep{ali2021spike2vec}}
          & SVM & 0.81 & 0.82 & 0.81 & 0.81 & 0.63 & 0.83 & 52.384 \\
        & NB & 0.65 & 0.77 & 0.65 & 0.64 & 0.46 & 0.74 & 9.031 \\
        & MLP & 0.81 & 0.82 & 0.81 & 0.81 & 0.52 & 0.77 & 44.982 \\
        & KNN & 0.80 & 0.80 & 0.80 & 0.79 & 0.54 & 0.75 & 2.417 \\
        & RF & 0.83 & 0.84 & 0.83 & 0.82 & 0.63 & 0.82 & 17.252 \\
        & LR & 0.82 & 0.84 & 0.82 & 0.82 & 0.68 & 0.83 & 48.826 \\
        & DT & 0.81 & 0.82 & 0.81 & 0.81 & 0.62 & 0.81 & 4.096 \\
        \midrule
        \multirow{7}{1.8cm}{PWM2Vec~\citep{ali2022pwm2vec}}
        & SVM & 0.83 & 0.82 & 0.83 & 0.82 & 0.68 & 0.83 & 40.55 \\
        & NB & 0.37 & 0.68 & 0.37 & 0.33 & 0.43 & 0.69 & 1.56 \\
        & MLP & 0.82 & 0.82 & 0.82 & 0.81 & 0.61 & 0.80 & 17.28 \\
        & KNN & 0.82 & 0.80 & 0.82 & 0.81 & 0.57 & 0.78 & 2.86 \\
        & RF & 0.84 & 0.84 & 0.84 & 0.84 & 0.69 & 0.83 & 5.44 \\
        & LR & 0.84 & 0.84 & 0.84 & 0.83 & 0.69 & 0.83 & 43.35 \\
        & DT & 0.82 & 0.81 & 0.82 & 0.81 & 0.62 & 0.82 & 3.46 \\
        \midrule
        \multirow{7}{1.8cm}{Approx. Kernel~\citep{farhan2017efficient}}
        & SVM & 0.78 & 0.79 & 0.78 & 0.77 & 0.53 & 0.78 & 16.67 \\
        & NB & 0.62 & 0.66 & 0.62 & 0.61 & 0.39 & 0.72 & 0.19 \\
        & MLP & 0.79 & 0.77 & 0.79 & 0.77 & 0.52 & 0.80 & 8.34 \\
        & KNN & 0.85 & 0.84 & 0.85 & 0.85 & 0.52 & 0.80 & 0.24 \\
        & RF & 0.82 & 0.81 & 0.82 & 0.81 & 0.54 & 0.83 & 1.95 \\
        & LR & 0.76 & 0.77 & 0.76 & 0.74 & 0.51 & 0.83 & 3.80 \\
        & DT & 0.77 & 0.77 & 0.77 & 0.77 & 0.47 & 0.82 & \textbf{0.27} \\
        \midrule
        \multirow{3}{2cm}{Neural Network}
        & LSTM & 0.33 & 0.10 & 0.33 & 0.16 & 0.02 & 0.5 & 16724.06 \\
        & CNN 
        & 0.26 & 0.11 & 0.26 & 0.13 & 0.08 & 0.55 & 41470.84 \\
         & GRU 
        & 0.31 & 0.10 & 0.31 & 0.15 & 0.03 & 0.5 & 444199.8 \\
        \midrule 
        \multirow{7}{1.8cm}{MFV}
          & SVM & 0.79 & 0.80 & 0.79 & 0.78 & 0.71 & 0.86 & 46.219 \\
        & NB & 0.48 & 0.69 & 0.48 & 0.46 & 0.59 & 0.74 & 7.206 \\
        & MLP & 0.76 & 0.79 & 0.76 & 0.76 & 0.57 & 0.78 & 101.491 \\
        & KNN & 0.71 & 0.82 & 0.71 & 0.75 & 0.63 & 0.82 & 2.157 \\
        & RF & 0.80 & 0.81 & 0.80 & 0.79 & 0.76 & 0.87 & 17.928 \\
        & LR & 0.80 & 0.81 & 0.80 & 0.79 & 0.77 & 0.87 & 8.631 \\
        & DT & 0.79 & 0.80 & 0.79 & 0.78 & 0.65 & 0.86 & 1.250 \\
        \midrule
        \multirow{7}{1.8cm}{PSWM2Vec}
          & SVM &  0.75 & 0.79 & 0.75 & 0.74 & 0.77 & 0.89 & 3.51 \\
        & NB & 0.59 & 0.66 & 0.59 & 0.57 & 0.53 & 0.80 & 0.22 \\
        & MLP & 0.78 & 0.79 & 0.78 & 0.76 & 0.76 & 0.87 & 8.61 \\
        & KNN & 0.77 & 0.78 & 0.77 & 0.77 & 0.68 & 0.84 & 1.39 \\
        & RF & 0.80 & 0.82 & 0.80 & 0.80 & 0.79 & 0.88 & 1.11 \\
        & LR & 0.75 & 0.76 & 0.75 & 0.73 & 0.70 & 0.85 & 1.54 \\
        & DT & 0.77 & 0.78 & 0.77 & 0.76 & 0.70 & 0.88 & \textbf{0.17} \\
        \midrule 
        \multirow{7}{2cm}{Virus2Vec}
          & SVM & 0.85 & 0.86 & 0.85 & 0.85 & 0.873 & 0.93 & 105.45 \\
        & NB & 0.67 & 0.78 & 0.67 & 0.65 & 0.65 & 0.83 & 3.98 \\
        & MLP & 0.85 & 0.85 & 0.85 & 0.84 & 0.79 & 0.90 & 32.68 \\
        & KNN &  0.84 & 0.85 & 0.84 & 0.83 & 0.76 & 0.88 & 54.72 \\
        & RF & 0.86 & 0.86 & 0.86 & 0.85 & 0.84 & 0.91 & 7.53 \\
        & LR & \textbf{0.87} & \textbf{0.87} & \textbf{0.87} & \textbf{0.86}& \textbf{0.877} & \textbf{0.94} & 5.57 \\
        & DT & 0.81 & 0.82 & 0.81 & 0.81 & 0.76 & 0.88 & 1.11 \\
        \bottomrule
    \end{tabular}
    }
    \caption{Performance comparing for different embedding methods on Aligned \textbf{Spike Sequence} data. The best values are shown in bold. The $\uparrow$ and $\downarrow$ mean higher and lower values are better, respectively.}
    \label{tbl_no_feat_select_results_aligned}
\end{table}

\section{Conclusion}\label{sec_conclusion}
We propose an efficient sequence embedding approach named Virus2Vec, which uses an alignment-free method based on minimizers and PWM to classify hosts of different coronaviruses using spike sequences. 
Virus2Vec not only performs better as compared to other methods requiring sequence alignment but also is better since it is an alignment-free approach.  
We show that our approach for unaligned sequences is efficient to generate compared to the popular alignment-free methods and has comparable predictive performance and better runtimes. 
In the future, we would focus on collecting more data to evaluate the scalability of Virus2Vec.  Such an approach could also work even on \emph{unassembled} (short read) data (not just unaligned), in a similar way that it works for metagenomics.


\bibliography{bibliography}

\end{document}